\documentclass{ametsocV6.1}
\usepackage{eso-pic}
\usepackage{xcolor}
\usepackage{latexml}

\usepackage{fancyhdr}
\usepackage{longtable}
\usepackage{array}
\usepackage{makecell}   
\usepackage{booktabs}   
\usepackage{float}      
\usepackage[section]{placeins} 
\makeatletter
\AtBeginDocument{%
  \nolinenumbers
}
\makeatother

\title{Tropical Cyclone Wind Field Diagnosis with Image-to-Image Models}

\authors{Sarah J. Ollier,\aff{a}\correspondingauthor{Sarah J. Ollier, sjo@worldsphere.ai} 
David S. Nolan,\aff{a,b}   
Brian D. McNoldy,\aff{a,b}   
Eduardo Siman,\aff{a,b}}  
\affiliation{\aff{a}{Worldsphere.ai}\\
\aff{b}{Rosenstiel School of Marine, Atmospheric, and Earth Sciences, University of Miami}\\
}

\abstract{
This study presents a novel application of image-to-image generative adversarial networks (GANs) to produce a complete tropical cyclone (TC) surface wind field diagnosis from a single frame of satellite infrared brightness temperature (IR). Because TC surface wind fields are infrequently observed, we train on two data sets. The first contains 3,323 image pairs of IR and wind fields from Weather Research and Forecasting Model (WRF) simulations (WRF2WRF). The second contains 11,886 real IR satellite images paired with parametric wind fields built on the extended best-track data set (IBTrACS) of TC intensities and wind radii (GOES2PARA). Both models, conditioned on storm intensity, motion, and latitude, transform single-channel IR into wind fields over 6$^{\circ} \times$ 6$^{\circ}$ (WRF2WRF) and 7$^{\circ} \times$ 7$^{\circ}$ (GOES2PARA) domains, at 256 $\times$ 256 pixel resolution. GOES2PARA captures outer wind field asymmetries, enabling diagnosis of 64, 50, and 34-knot wind radii (R64, R50, R34) in each quadrant. WRF2WRF also produces detailed rainbands and eyewall asymmetries. Pixel-wise RMSE against held out truth demonstrates realistic wind field generation with mean RMSE of 6.4 knots (GOES2PARA) and 7.2 knots (WRF2WRF). Compared against IBTrACS, WRF2WRF achieved mean absolute errors (MAE) of 14.8 nautical miles (NM) for radius of maximum wind (RMW) and 25.9 NM for R34. GOES2PARA achieved 10.0 NM MAE for RMW and 25.9 NM for R34. Against Synthetic Aperture Radar wind fields GOES2PARA yielded a lower mean pixel-wise error (RMSE 12.1 kt, MAE 9.5 kt) than WRF2WRF (RMSE 13.9 kt, MAE 11.3 kt). Unlike many existing operational surface wind speed algorithms, our image-to-image models provide updated wind fields with every new satellite image.}

\begin{document}
\maketitle

\iflatexml
\begin{center}
\color{gray}\small
Preprint prior to peer-review. This Work was submitted to Artificial Intelligence for the Earth Systems on 8th September 2026. Copyright in this Work may be transferred without further notice. Copyright in this Work may be transferred without further notice. Generated using the official AMS \LaTeX\ template.
\end{center}
\fi

\statement 
Knowing the size and shape of a tropical cyclone (TC) is essential for understanding and forecasting its impacts. However, tropical cyclone wind fields are rarely measured. We present two computer vision models that reconstruct two-dimensional TC surface wind fields, including storm asymmetry, from a single operational satellite image and widely available storm parameters. These models can produce a TC wind field from a satellite image in less than 1 second and as frequently as satellite data is available. When tested against current methods and radar derived winds, they performed comparably, suggesting they could support near real-time monitoring of tropical cyclone structure.

\section{Introduction}
\label{intro}

In coordination with the World Meteorological Organization (WMO), the 6 Tropical Cyclone Regional Specialized Meteorological Centers (TC-RSMCs) and the 4 Tropical Cyclone Warning Centers (TCWC) share responsibility for providing analyses and forecasts for tropical cyclones around the world \citep{WMO2026}. In addition to position, motion, and intensity (estimated maximum surface winds), the centers also provide estimates and forecasts of the areas of tropical-storm force and hurricane-force winds for each of the northeast (NE), southeast (SE), southwest (SW) and northwest (NW) quadrants around the center \citep{Knaff2017, DeMaria_2022}. In the North Atlantic Ocean, and sometimes in the East Pacific, the United States executes aircraft reconnaissance for most storms that could possibly make landfall, such that about 30\% of all wind intensity and size estimates have the benefit of in-situ measurements \citep{Rappaport_2009}. Nonetheless, without aircraft observations, or the (very) occasionally well-situated island or ship report, the large majority of intensity analyses in the North Atlantic and virtually all analyses around the rest of the world must be made from infrared (IR) satellite images from the geostationary satellites. For many years, the subjective (human) pattern recognition scheme developed by Dvorak (\citeyear{Dvorak1984}; see also \citealp{Velden2006}) was the method used by most operational centers. More recently, many objective (computer analysis) methods have been developed and are now part of a suite of algorithms that estimate TC intensity from IR images \citep{Olander_2007, Olander2019, Ritchie_2012, VELDEN2025}.  

While it is generally believed that TC intensity is quite accurately assessed for every synoptic analysis, the size of the wind field (as described by the so called wind-radii in each quadrant) has less consistent observational support. Satellite based scatterometers (active radars that convert ocean surface roughness into wind speeds) can provide highly useful images of the surface wind field, thus supplying critical information for the quadrant wind radii estimates \citep{Brennan_2009_scat, Knaff2011, knaff_2021}. However, being low-orbit satellites, the scatterometer foot prints are only a few hundred kilometers wide and return times vary from 12 to 24 hours. Thus their views of the TC wind field are relatively infrequent and often incomplete due to partial coverage. Furthermore, scatterometer winds become increasingly uncertain at high wind speeds with full signal saturation at 30-35~ms$^{-1}$ (60-70 kt) depending on radar band, and winds above $\sim$25~ms$^{-1}$ ($\sim$50 kt) also showing signs of degradation \citep{Wentz_1999, Verspeek_2010, Ricciardulli2015, Chou_2013_ascat, Soisuvarn_2013, Chiara_2016_ascat, Stoffelen_2017, knaff_2021}. Rain contamination further degrades wind retrieval in the TC inner core \citep{Chou_2013_ascat, Chiara_2016_ascat}, whilst steep wind gradients near the RMW also increase wind retrieval errors \citep{knaff_2021}.

A different remote sensing technique that can be useful in determining critical wind radii in tropical cyclones is Global Navigation Satellite Systems reflectometry, or GNSS-R. The principal behind this is a bi-static radar, where one satellite is simply a receiver of "signals of opportunity" that reflect off the ocean surface from a transmitting GPS satellite. CYGNSS (CY for cyclone) has been a pioneering mission in this, and has three important advantages over conventional scatterometry: 1) a low-inclination orbit allows for more frequent revisit times over TCs, 2) the L-band (19 cm) GPS signals do not attenuate in precipitation, and 3) the GMF for these types of signals saturates at a much higher wind speed threshold (70~ms$^{-1}$, 136 kt; \citealp{Clarizia_2016_cygnss, knaff_2021}). Although there are inherent limitations in this technique as well (such as irregular data spacing and noisy retrievals), it can provide skillful analyses of R34 and R50 at approximately 50\% of all best-track times \citep{McNoldy2026}. 

Wind retrieval from the sea-surface roughness measured by Synthetic Aperture Radar (SAR) Normalized Radar Cross Section (NCRS) provides instantaneous snapshots of kilometer-resolution ocean surface wind fields. These highly detailed surface wind fields allow for calculation of TC structural parameters, including sustained maximum wind speed ($V_{max}$), Radius of Maximum Wind (RMW) and quadrant wind radii \citep{Mouche_2019, Combot2021_SAR_TC}. Extreme rainfall remains a challenge for SAR wind retrieval due to signal degradation \citep{Mouche2026_TC_SAR}. SAR retrievals are available at higher wind speeds than scatterometers (up to $\sim$75 ~ms$^{-1}$, $\sim$150 kt) \citep{Mouche_2019}. Similarly to scatterometry, SAR wind field retrieval relies on the satellite overpass coinciding with the storm location. Despite improvements in SAR TC observation frequency in recent years, wind swath availability remains low with approximately 6-8 observations per observed TC during 2023-2025 and approximately 16 percent of TCs with no observations in 2025 \citep{Mouche2026_TC_SAR}. Furthermore, SAR wind swaths are 350-500 kilometers wide and are not routinely centered on the storm, frequently leading to partial coverage. For more details on SAR based wind retrieval for TCs, the reader is referred to \citet{Mouche2026_TC_SAR}. A comprehensive review of all the instruments and algorithms for estimating TC size is provided in \citet{knaff_2021}.

Whilst the direct TC wind field measurements discussed above remain infrequent, saturate at high wind speeds, and are spatially incomplete, geostationary passive IR satellite imagery provides near-continuous high resolution global coverage and is freely and operationally available in near real-time. For example, hourly Geostationary Operational Environmental Satellite (GOES) provides a consistent record spanning the full life cycle of current and historical TCs \citep{goes_gridsat, noaa_goes_aws}. This vast and growing dataset already underpins operational intensity and size analyses such as MTCSWA \citep{knaff_2021}, yet it remains under-utilized for reconstruction of complete, spatially continuous surface wind fields. In recent years, increasing contributions to TC wind diagnosis, including from IR, have been based on advances in deep learning methodologies. 

Deep learning advances include TC intensity diagnosis \citep{Pradhan_2018, Chen2019, yang_2024, ZhuoTan2021, xu_tropical_2023}, short term intensity changes \citep{Griffin_ST_intensity_2024_cnn}, and TC wind radii \citep{Wang2023, ZhuoTan2021} with Convolutional Neural Networks (CNNs), and inner core wind field estimation \citep{Wimmers2024} with Unets. More recently there have been developments in 2-dimensional TC wind field generation with Generative Adversarial Networks (GANs) \citep{Ollier_2025, cheng_2026_GAN} and 3-dimensional wind fields with diffusion models \citep{han_diffusion_3D_2026}. Many of these deep learning studies use multiple input datasets from a variety of observational and reanalysis data sources, all of which have inherent biases. For example, \citet{cheng_2026_GAN} utilizes IR, synthetic passive microwave and gridded reanalysis surface wind data for 2-dimensional TC surface wind asymmetry analysis. Lighter-weight training datasets for two-dimensional TC surface-wind diagnosis include geostationary IR TC imagery paired with best-track wind radii \citep{Wang2023}, and IR paired with parametric wind fields or with wind-field simulation data \citep{Ollier_2025}. Whilst \citet{Wang2023} labeled satellite images with per-quadrant scalar radii to produce scalar model outputs, \citet{Ollier_2025} employed purely image-to-image models to produce whole storm wind fields.

This study builds on recent developments in deep-learning retrieval of TC structural information from satellite imagery to exploit the near-continuous, freely available IR record as a proxy for the wind observations that are infrequently available during active storms. We present two parameter conditioned image-to-image machine learning models that translate near-continuous satellite IR imagery into complete, surface wind fields spanning the storm center to outer rain bands. The two models are trained on two distinct data sets. WRF2WRF is trained on entirely synthetic data and is derived from model output from 24 different simulations of TCs, which provide structural detail difficult to obtain consistently from observations. GOES2PARA model utilizes satellite IR data paired with wind fields from a novel parametric model with operational best track data inputs. 

The remainder of this paper is organized as follows: Section 2 introduces the datasets and methodology for obtaining high resolution TC wind fields using WRF2WRF and GOES2PARA image-to-image models. Model results, benchmarking and placement within the broader literature are discussed in Section 3. Conclusions and future research directions are provided in Section 4.

\section{Data and Methodology}
\label{sect:data_methods}

\subsection{WRF simulations}
\label{sect:wrf_methods}

The WRF2WRF model is trained on simulated satellite images and surface wind fields from 24 simulations of hurricanes produced by the Weather Research and Forecasting Model (WRF). These are collected from simulations produced for several different research projects dating back to the Hurricane Nature Run (HNR) produced by \citet{Nolan2013}. The HNR was later supplemented with four additional nature, idealized, and real-case simulations as described in \citet{Klotz_Nolan_2019}. More recently, 19 additional simulations of real hurricanes from the 2020, 2023, 2024 and 2025 seasons were added to the set. Information about all 24 simulations, including the source of the initial and boundary condition data, are listed in Table A.1 in the Appendix~\ref{app:wrf}.

Although the versions of the WRF model have advanced over the years from 3.1.1 to 4.3.3, all the simulations here used the same parameterizations, the same vertical levels, and grid spacings, and similar configurations of domain and nested grid sizes. In particular, most of the simulations use an outer domain (D01) with 27 km grid spacing, and 3 vortex-following nested grids with 9 km (D02), 3 km (D03), and 1 km (D04) spacing; a few simulations do not use the 27 km grid so that D01 is 9 km, etc. Most of the simulations use 480x480 grid points on D04, but some use 360x360. All use 240x240 or 360x360 points on D03, which is the grid that is used to produce the synthetic infrared and wind fields. The time steps, parameterizations, additional aspects of the simulations, and a list of storms are provided in the Appendix~\ref{app:wrf}. 

Model output from D03 at hourly intervals was used to produce simulated IR images and surface winds. The RRTM-G longwave radiation scheme (Iacono et al. 2008) produces outgoing longwave radiation (OLR) as an output variable, but it does not provide IR brightness temperature ($T_B$). Therefore, $T_B$ is estimated from from a simple linear formula as suggested in Vaillant de Guelis et al. (2017) and Christensen and Driver (2021):
\begin{equation}
  T_B = \frac{\mathrm{OLR}+b}{a}
\end{equation}
		
The aforementioned papers recommend $a = 2.0 ~W m^{-2} K^-1$ and $b = 310 ~W m^{-2}$. By comparing to GOES IR images from real hurricanes of similar intensities, we found that, using this formula, the resulting $T_B$ values were generally too warm for the highest clouds by about 10 K. For a better correspondence with real images, we use $b = 290 ~W m^{-2}$ instead.

Most of the simulated storms had little or no interaction with land. However, for those that passed near or over land, the surface wind speed over the land areas was adjusted using the logarithmic law to extrapolate the lowest model level wind speed down to 10 m height over a surface with roughness length $z_0$ = 0.003 m.

\subsection{Geostationary Satellite TC cloud brightness temperature}
\label{sect:goes_methods}

For GOES2PARA we utilized cloud brightness temperature from 
Geostationary Operational Environmental Satellite (GOES) imagery. Hourly images for storms between 2017--2026 were obtained from  GOES-16 and GOES-19 
infrared (IR) band~13 (${\sim}$10.3~$\mu$m) Advanced Baseline Imager 
Level-2 Cloud and Moisture Imagery (ABI-L2-CMIP), available from the 
NOAA AWS Open Data Registry \citep{noaa_goes_aws} at 2~km resolution. 
For older storms between 2005--2017, we utilized hourly GOES-12 and GOES-13 
IR channel~4 (${\sim}$10.7~$\mu$m) from the Gridded Satellite 
(GridSat) GOES dataset \citep{goes_gridsat} at 4~km resolution.

\subsection{Storm size data}
\label{sect:ibtracs_methods}

Operational best track data from International Best Track Archive for Climate Stewardship (IBTrACS; \citep{Knapp2010_ibtracs1,Gahtan2024_ibtracs2}) provided storm Radius of Maximum Wind (RMW) and quadrant wind radii (34, 50, and 64 kt) for benchmarking model-derived wind structure. For comparison against best track, GOES IR images were supplied as input for inference for both WRF2WRF and GOES2PARA. Wind radii were derived from model output by first applying a center finding algorithm to the generated wind field, where the storm center was refined to the wind-speed minimum within 50 km of the image center. The RMW was computed as the radius of the azimuthal-mean ($360^\circ$) tangential wind maximum. Quadrant radii were computed by azimuthally averaging the wind field within each 90° quadrant and identifying the outermost radius at which the quadrant-mean wind met or exceeded each threshold. We applied a radial sampling grid extending to $(nr - 0.5)\times \Delta r$ from the storm center, where $\Delta r$ is the pixel spacing for model's input domain (6$^{\circ}$ for WRF2WRF, 7$^{\circ}$ for GOES2PARA, both at 256 $\times$ 256 pixel resolution); $nr$ was set per model 110 for WRF2WRF ($\simeq$ 154.6 NM) and 128 for GOES2PARA ($\simeq$ 210 NM, the maximum supported by its coarser 7$^{\circ}$ pixel spacing). Whilst operational centers use maximum, rather than mean, for per quadrant radii calculation, in preliminary testing with WRF2WRF we found calculating maximum extent lead to higher errors than mean for R34. We attribute this larger error to pixel noise in the outer rain bands. Wind speed was recovered from the model imagery by nearest-neighbor lookup against the colormap (256 levels, 0--100 kt). Quadrant comparisons were restricted to cases where both model and best track radii fell within the 6$^{\circ}$ domain for WRF2WRF and 7$^{\circ}$ domain for GOES2PARA; the excluded fraction was reported per threshold. Systematic offsets exist between model-derived radii and best track estimates, thus a bias correction was applied prior to evaluation. For each wind speed threshold (34, 50, and 64 kt), bias (model minus IBTrACS) was computed per quadrant over the validation set. This bias was then used to correct the corresponding model-dervied radii from test-set inference prior to comparison with IBTrACS.

\subsection{Parametric TC wind fields}
\label{sect:para_methods}

Two-dimensional wind fields were generated using a novel parametric model developed to approximate TC surface wind fields. The size, strength, and asymmetries of the wind field are derived from the IBTraCS dataset. The storm parameters from 6-hourly synoptic data were interpolated in time to create smooth 1-hourly wind fields on a polar coordinate grid (radial spacing 4 km, angular spacing $2^\circ$) with translation asymmetry following \cite{Knaff2007} and size asymmetry controlled by tropical storm wind radii in four quadrants. Wind fields at each time step were then interpolated onto a Cartesian coordinate grid with 2 km resolution. The model is described in more detail in Appendix~\ref{app:parametric}. We selected 54 Atlantic basin storm cases spanning 2004-2025, with peak intensities ranging from tropical storm to Category 5 hurricane. The beginning of the dataset in 2004 corresponds to when tropical storm wind radii (R34) were carefully reanalyzed. Only open-ocean cases were selected to avoid the confounding effect land can have on the infrared satellite presentation and its relationship to the surface wind field. For storms that approached land or made landfall, only the portion of their track over water was used.

\subsection{Synthetic Aperture Radar Wind Fields}
\label{sect:sar_methods}

To benchmark whole wind-field quality, model inferences were compared against Synthetic Aperture Radar (SAR) Tropical Cyclone Wind products at 3~km resolution from NOAA's STAR Water Surface Conditions Science Team’s tropical cyclone winds program \citep{NOAA_STAR_SAR}. Model inference on hourly GOES imagery were matched to concurrent SAR scenes of the same storm within a maximum time difference of 30~minutes. This yielded 42 SAR scenes from 5 unique storms for WRF2WRF and 28 from 5 unique storms for GOES2PARA. To evaluate the parametric model the same 28 SAR scenes were compared with the parametric wind fields from the GOES2PARA test set. The sample sizes per storm vary as older storms present in the test sets coincided with few to no SAR satellite passes, whilst more frequent SAR imagery was available for more recent storms. We excluded SAR land pixels, non-retrieval pixels and swath-edge pixel artefacts which arise from antenna roll-off and backscatter contamination. Model storm center was refined using the same storm-centering methodology described in Section 2\ref{sect:ibtracs_methods}. The model wind field was then interpolated bi-linearly onto the native SAR grid. Pixel-wise model vs SAR RMSE, MAE and bias were computed over the intersection of valid pixels in both fields. 

\subsection{Image processing}
\label{sect:img_processing}

\begin{figure}[H]
  \centerline{\includegraphics[width=29pc]{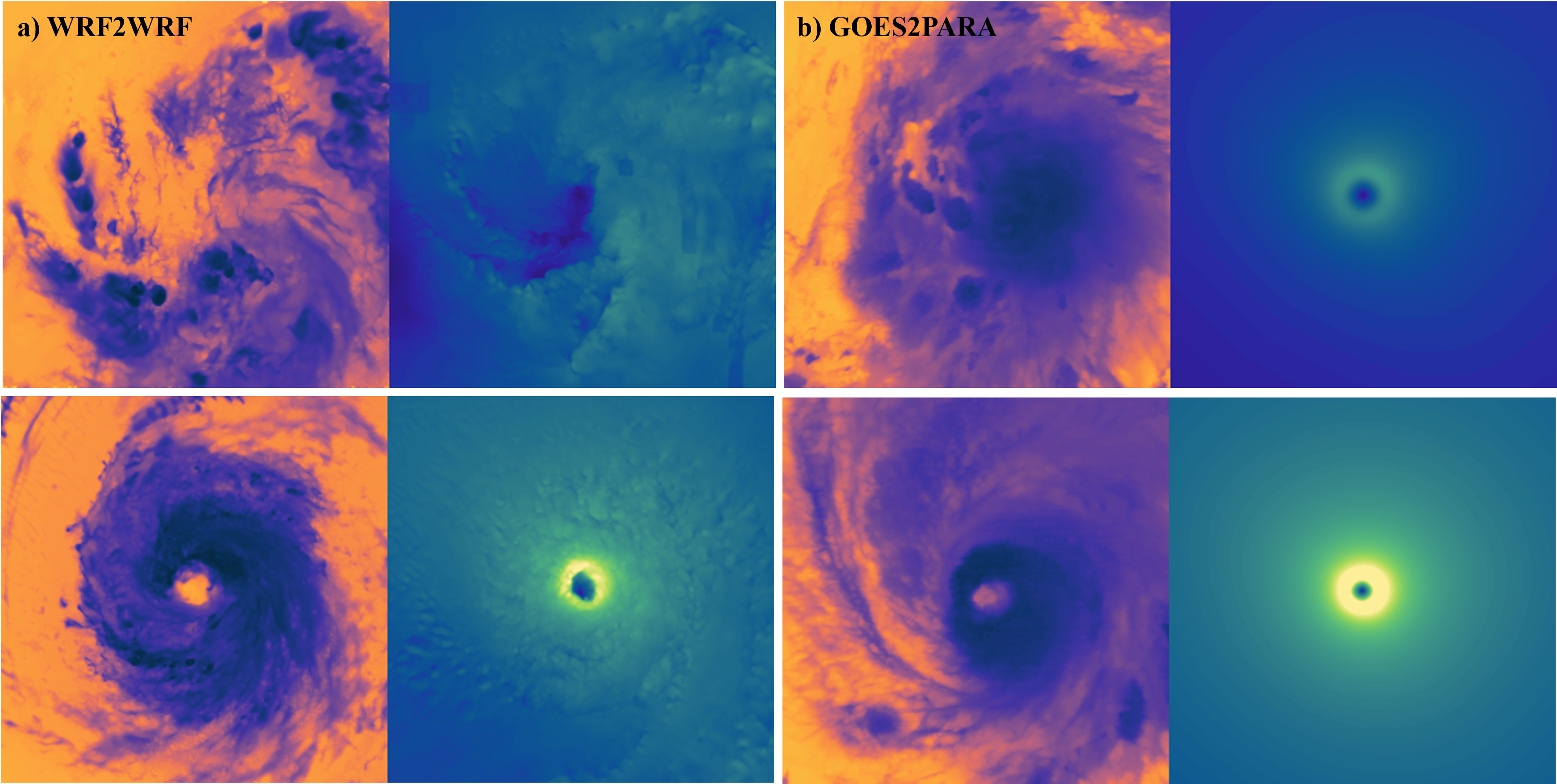}}
\caption{Example training data images: (a) WRF2WRF: Ernesto, Bill. (b) GOES2PARA: Jose (AL122017), Ivan (AL12004).}
  \label{input_data}
\end{figure}

WRF2WRF data were plot as $6^{\circ} \times 6^{\circ}$ images with 256 $\times$ 256 pixels centered around the storm eye. GOES2PARA data were plot as $7^{\circ} \times 7^{\circ}$ images with 256 $\times$ 256 pixels. Whilst WRF2WRF IR and wind field were obtained from WRF similution data (Section 2\ref{sect:wrf_methods}), GOES2PARA pairs GOES-IR images with concurrent parametric wind fields (Sections 2\ref{sect:goes_methods}, 2\ref{sect:para_methods}). WRF2WRF and GOES2PARA wind field images were rendered with the perceptually uniform cmocean haline colormap \citep{Thyng2016}. The color range spanned from 0 to 100 knots with 100 contour levels. The wind speed was saturated at 100 knots to allow for more color variations across the lower wind speeds which are important for diagnosing TC size. WRF2WRF and GOES IR images were plot using cmocean thermal colormap \citep{Thyng2016}, with the color range between -90-40 $^{\circ}$C with 130 contour levels.  The contour-level counts were chosen so that each level spans a comparable increment in the underlying variable: 1 kt per level for wind and 1$^{\circ}$C per level for brightness temperature, whilst value ranges were selected to emphasize key meteorological thresholds in the data based on preliminary testing. Perceptually uniform colormaps were selected to represent the continuous data rather than scalar values, providing a consistent three-channel representation for the image-translation architecture. Because the models operate on images rather than on data matrices, the color mappings chosen for the IR and surface wind fields may influence the quality of the results. Although each grid cell corresponds to a single brightness-temperature or wind-speed value, and could therefore be represented in a single grayscale channel, early experiments found that mapping to three-channel color imagery improved results relative to grayscale. For both models concurrent IR and wind field images were concatenated into $512 \times 256$ pixel images as shown in Figure~\ref{input_data}.

\subsection{Train, Validation, Test sets}
\label{sect:train-val-test}

All datasets were categorized by Saffir-Simpson Hurricane Wind Scale per storm and per image before creation of the train, test, and validation sets with approximately 70:15:15 split ratios (67:17:16 for WRF2WRF, 70:15:15 for GOES2PARA). To prevent data leakage, splits were performed at the storm level such that all images from a given storm appear exclusively in one set. Storms were assigned to sets aiming for a representative spread of storm categories and individual image sustained maximum wind speed ($V_{\text{max}}$) across all three sets. Train, validation and test dataset sizes are included in Table ~\ref{tab:app_gan_parameters}. 

In addition to validation and testing of WRF2WRF with simulated storm data from the WRF2WRF validation and test sets (Section 3\ref{sect:res_model_test_set}), we assess the WRF2WRF performance on real satellite imagery as no real storm data was included in WRF2WRF training and validation sets. We used $6^{\circ} \times 6^{\circ}$ $256 \times 256$ pixel GOES-IR imagery for final model inference to replace $6^{\circ} \times 6^{\circ}$ $256 \times 256$ pixel WRF simulation data. No truth image is required for final model inference.

We replaced WRF2WRF validation and test set with real GOES-IR data for storms that were not in the WRF2WRF training or validation sets, yielding 1661 GOES-IR validation images from 7 storms and 1593 GOES-IR test images from 7 storms, close to the validation and test set size for GOES2PARA (Table~\ref{tab:app_gan_parameters}). 

The same test and validation sets were not used for both models as some storms in the test and validation sets for GOES2PARA featured as simulatied storms in the WRF2WRF training or validation data. The WRF2WRF model was not retrained or fine-tuned for this step providing a zero-shot transfer test of the domain shift from synthetic IR to GOES-IR. In the results sections we refer to this test as WRF2WRF(GOES) to distinguish from the purely simulation data results for WRF2WRF.

\subsection{Model architecture, software and hardware}
\label{sect:model_arch_software_hardware_methods}

\begin{figure}[H]
  \centerline{\includegraphics[width=38pc]{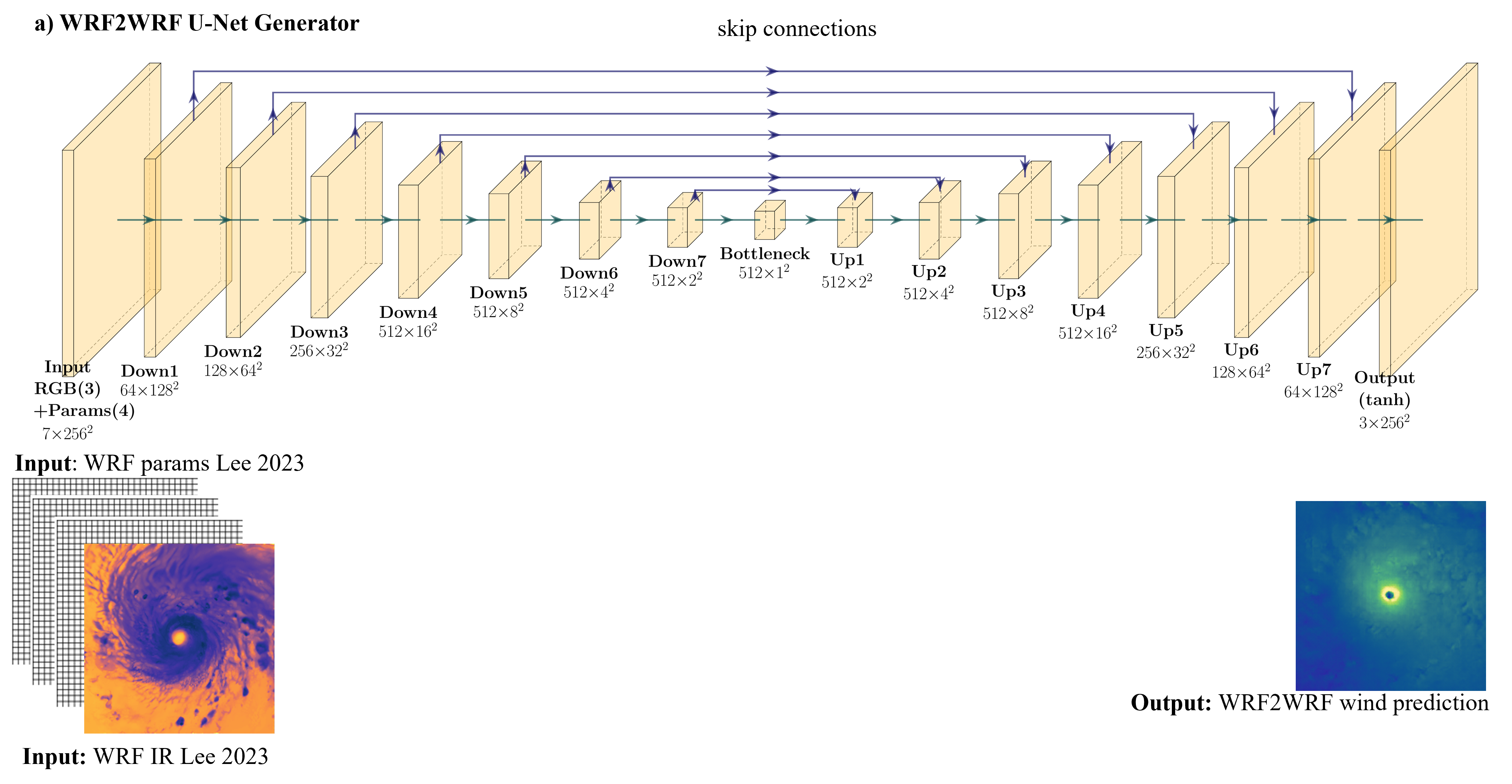}}
  \centerline{\includegraphics[width=38pc]{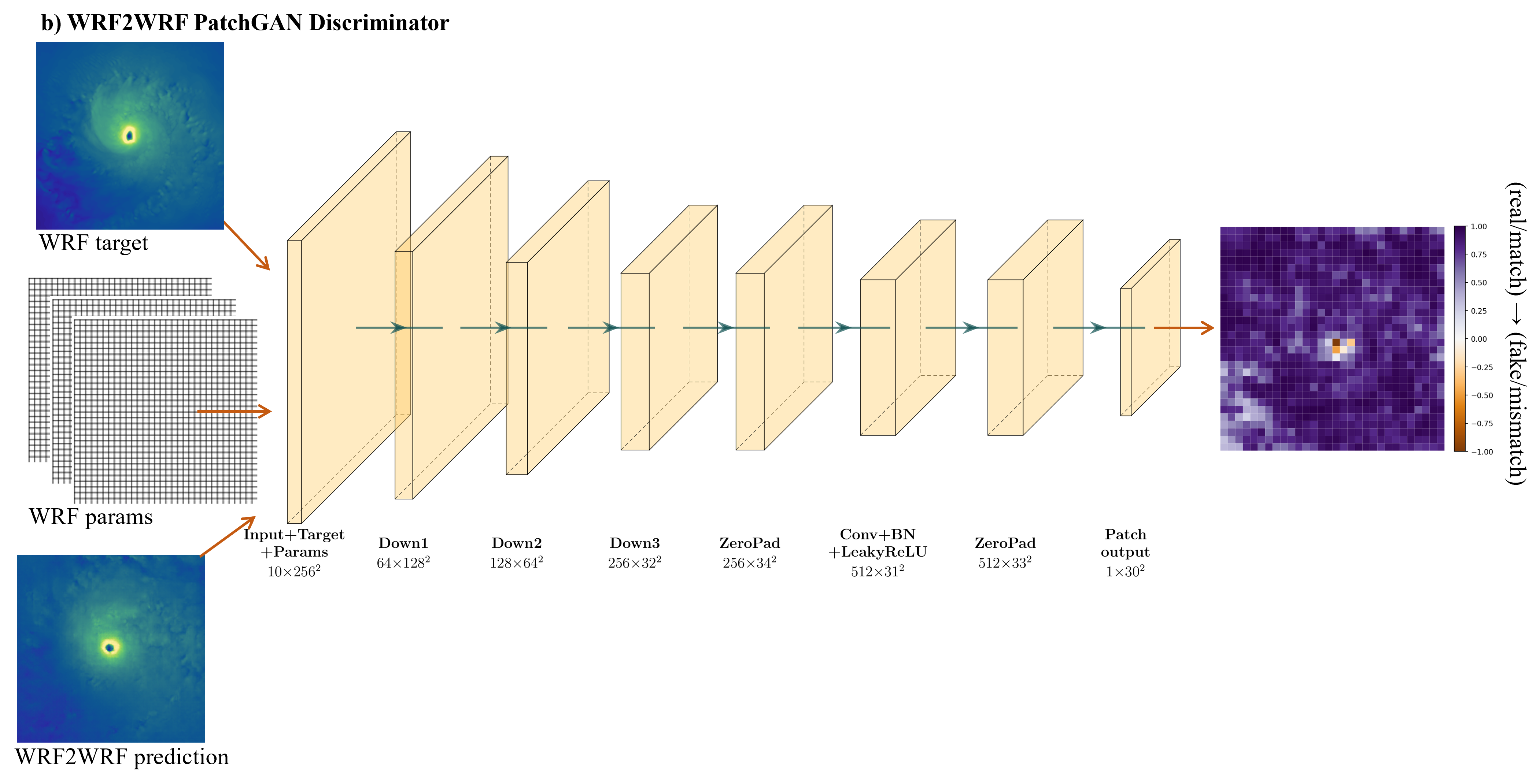}}
  \caption{Parameter conditioned Pix2Pix GAN \citep{Isola2017}. Parameters: Coriolis proxy: sin($\phi$), sustained maximum wind ($V_{\text{max}}$) and motion vectors ($u$, $v$) are normalized and broadcast spatially to a 256×256×1 field and concatenated with the input image channels to produce a 7 channel concatenated input of image Red, Blue, Green (RGB, 3), $V_{\text{max}}$(1) , sin($\phi$)(1) and ($u$, $v$)(2). (a) The input data is encoded and decoded via the U-Net generator to produce a generated storm image. (b)The Patch GAN Discriminator then predicts whether the generated and target image patches are real or fake in patches across the input and generated image and averages this to a real or fake label. The generator is skilled when the discriminator is unable to detect fake images, labeling them as real. The image examples show WRF2WRF, GOES2PARA uses the same architecture with GOES2PARA data. For ablation studies the omitted conditioning parameters were set to 0 to preserve the same architecture throughout.}
  \label{pix2pix}
\end{figure}

WRF2WRF and GOES2PARA both utilise a modified Pix2Pix GAN architecture  \citep{Isola2017}. The number of model layers follows \cite{Isola2017} with modifications to the Generator and Discriminator (Fig.~\ref{pix2pix}) to condition the input images with storm parameters. WRF2WRF and GOES2PARA were conditioned on three physical parameters that are routinely available in real-time operational advisories: storm-center latitude, maximum sustained wind speed ($V_{\text{max}}$), and the storm motion vector ($u$, $v$). Latitude was encoded as $\sin(\phi)$, a monotonic proxy for the Coriolis parameter $f = 2\Omega\sin(\phi)$ , where $\Omega$ is the
Earth's rotation rate, allowing the network to learn the appropriate scaling and sign relationship between latitude and background rotational effects on storm structure. Each scalar (or two-component vector, in the case of motion) parameter
was passed through a small dedicated encoder network. The three conditioning parameters were mapped to normalized ranges, spanning the dataset ranges, prior to spatial broadcasting. The encoding of parameters in the models follows:

\paragraph{Latitude (Coriolis proxy).} Storm-center latitude $\phi$ was transformed via $\sin(\phi)$ and normalized to $[0,1]$ using fixed reference bounds $\phi_{\min}$ and
$\phi_{\max}$:
\begin{equation}
  \widetilde{\phi} = \mathrm{clip}\!\left(
    \frac{\sin(\phi) - \sin(\phi_{\min})}{\sin(\phi_{\max}) - \sin(\phi_{\min})},\;
    0,\,1
  \right)
  \label{eq:lat_encoder}
\end{equation}
with $\phi_{\min} = 7^{\circ}$ and $\phi_{\max} = 40^{\circ}$ for WRF2WRF and GOES2PARA.
\paragraph{Maximum sustained wind speed.} $V_{\text{max}}$ (kt) was
normalized to $[0,1]$ using fixed reference bounds $V^{\text{ref}}_{\min}$
and $V^{\text{ref}}_{\max}$:
\begin{equation}
  \widetilde{V}_{\text{max}} = \mathrm{clip}\!\left(
    \frac{V_{\text{max}} - V^{\text{ref}}_{\min}}{V^{\text{ref}}_{\max} - V^{\text{ref}}_{\min}},\;
    0,\,1
  \right)
  \label{eq:vmax_encoder}
\end{equation}
with $(V^{\text{ref}}_{\min}, V^{\text{ref}}_{\max}) = (30, 160)$~kt for WRF2WRF and $(34, 165)$~kt for GOES2PARA.
\paragraph{Storm motion vector.}
The storm motion components $u$, $v$ (km~h$^{-1}$) were
each normalized symmetrically about zero using a common bound
$M_{\text{ref}}$:
\begin{equation}
  \widetilde{u} = \mathrm{clip}\!\left(\frac{u}{M_{\text{ref}}},\,-1,\,1\right),
  \qquad
  \widetilde{v} = \mathrm{clip}\!\left(\frac{v}{M_{\text{ref}}},\,-1,\,1\right)
  \label{eq:motion_encoder}
\end{equation}
with $M_{\text{ref}} = 46$~km~h$^{-1}$ for WRF2WRF and $M_{\text{ref}} =28$~km~h$^{-1}$ for GOES2PARA due to differences in their sample storm motion ranges.

Each resulting scalar ($\widetilde\phi$, $\widetilde{V}_{\text{max}}$) or
two-component vector ($\widetilde u, \widetilde v$) was then broadcast uniformly across a $256\times256$ spatial grid into constant-value tensors. The resulting tensors were concatenated channel-wise with the three-channel input image (red--green--blue; RGB), yielding a 7-channel tensor (3: image RGB + 1 $\sin(\phi)$ + 1 $V_{\text{max}}$ + 2 ($u$, $v$)) then supplied to the U-Net encoder--decoder generator (Fig. ~\ref{pix2pix}). 

WRF2WRF and GOES2PARA full models were conditioned on all four parameters; single-parameter ablation variants were trained by disabling all other parameter conditioning inputs (Table~\ref{tab:model_variants}). Unused parameter fields were not removed from the input tensor but instead multiplied by a binary mask (0 or 1) prior to concatenation. This preserves a constant input channel dimensionality for all models, so that performance differences between variants reflect the influence of the conditioning parameters rather than architectural differences. Table~\ref{tab:app_gan_parameters} shows a summary of parameters and hyper-parameters for the models and Table~\ref{tab:model_variants} shows the conditioning parameters used for the main models (WRF2WRF, GOES2PARA) and the ablation versions. 

Model training was completed on  NVIDIA L4 GPU with TensorFlow 2.20.0. Final model inference timing was benchmarked on Google Colab using a CPU (High-RAM) runtime (Intel Xeon CPU at 2.20GHz (4 physical cores / 8 threads)) with TensorFlow 2.20.0, processing images at a batch size of 1 to simulate live, single-frame GOES image availability.

\begin{table}[h!]
\caption{Pix2Pix GAN datasets and parameters for WRF2WRF and GOES2PARA models. WRF2WRF(GOES) does not include model parameters due to the original WRF2WRF model remaining unchanged for validation and testing with GOES-IR imagery.}
\label{tab:app_gan_parameters}
\begin{center}
\footnotesize
\begin{tabular}{lccc}
\topline

\textbf{Parameters} & \textbf{WRF2WRF} & \textbf{GOES2PARA} &\textbf{WRF2WRF(GOES)}\\
\midline
Dataset & synthetic IR and wind field & GOES IR and parametric wind field & GOES IR and parametric wind field \\ 
Batch\_size & 1 & 8 & - \\
dataset dates & 2005-2025 &  2005-2025 & 2008 - 2025\\
no. storms & 24 & 54 & - \\ 
conditioning params & $\tilde\phi$, $\widetilde{V}_{\text{max}}$, 
($\tilde u, \tilde v$) & $\tilde\phi$, $\widetilde{V}_{\text{max}}$, 
($\tilde u, \tilde v$) & -\\
no. images & 3,323 & 11,886 & - \\
Train dataset size & 2,229 (67 \%)& 8,319 (70 \%) & -\\ 
Train dataset storms & 15 & 38 & - \\
Validation dataset size & 537 (17 \%) & 1,754 (15 \%) & 1661 \\
Validation dataset storms & 5 &  7 & 8 \\
Test dataset size & 557 (16 \%) & 1813 (15 \%) & 1593 \\ 
Test dataset storms & 4 & 8 & 7 \\
Buffer size & Train dataset size & Train dataset size & - \\
learning rate & 2e-4 & 2e-4 & - \\
Optimizer & Adam & Adam & -\\
$\lambda$ & 100 & 100 & -\\
Total epochs & 160 & 130 & -\\ 
Storm basin & North Atlantic & North Atlantic & North Atlantic \\
Latitude range (°) & 9-38 & 8 - 40 & - \\
Vmax range (kt) & - & 34 - 160 & -\\
Vmax range (ms-1) & - & 18 - 85 & - \\

\botline
\end{tabular}
\end{center}
\end{table}

\begin{table}[t!]
\caption{Model variants and their embedded parameters for WRF2WRF and GOES2PARA.}
\label{tab:model_variants}
\begin{center}
\footnotesize
\begin{tabular}{llll}
\topline
\textbf{Model} & \textbf{Variant} & \textbf{Embedded Parameters} & \textbf{Tag} \\
\midline
WRF2WRF & all  & $\tilde\phi$, $\widetilde{V}_{\text{max}}$, 
($\tilde u, \tilde v$) & WRF2WRF  \\
WRF2WRF & lat  & $\tilde\phi$                      & WRF2WRF(lat)  \\
WRF2WRF & Vmax & $\widetilde{V}_{\text{max}}$                                 & WRF2WRF(Vmax) \\
WRF2WRF & uv   & ($\tilde u, \tilde v$)                                   & WRF2WRF(uv)   \\
WRF2WRF & base & -                                      & WRF2WRF(base) \\
\midline
GOES2PARA & all  & $\tilde\phi$, $\widetilde{V}_{\text{max}}$, 
($\tilde u, \tilde v$) & GOES2PARA  \\
\botline
\end{tabular}
\end{center}
\end{table}

\subsection{Model evaluation}
\label{sect:evaluation_methods}

Validation set images were used to assess model performance throughout training. Final models were selected on the basis of validation image quality; qualitatively through visual comparison of predicted and target fields and inspection for hallucinated storm structures and quantitatively, using pixel-wise RMSE, the structural similarity index (SSIM), and peak signal-to-noise ratio (PSNR). The selected models were then evaluated on the held-out test set using the corresponding conditioning parameters.

Generated images from the test and validation set were evaluated with pixel-wise RMSE where the pixel values were converted back to numerical values using the image colormap. Pixel numerical value in the generated image was compared with the corresponding pixel in the truth image. For benchmarking with IBTrACS (\citealp{Knapp2010_ibtracs1, Gahtan2024_ibtracs2}; Section 2\ref{sect:ibtracs_methods}) and SAR data (Section 2\ref{sect:sar_methods}), GOES IR images and best track parameters were used as input data for inference for both models. Pixel-wise RMSE was calculated between model prediction and model truth and SAR windfield data by computing wind speed from the RGB imagery by nearest-neighbour lookup against the cmocean haline colormap \citep{Thyng2016} (256 levels, 0–100 kt). 

\subsection{Assistive Use of Large Language Models} 

Portions of the GAN model code and data analysis code were developed, debugged and reviewed with the assistance of Claude Sonnet 3.0--4.5 and Opus 4.5--4.8 (Anthropic) and Chat-GPT 3.0--4.o (OpenAI) which were also used to assist translation between coding languages. Limited inline code completions were provided by VS Code Copilot integration with the Github default model. 

The same Claude models and Overleaf AI assistant (Digital Science) were used to draft some short manuscript sections to support organization of author ideas prior to author rewrite, to generate LaTeX commands, to review spelling and grammar, check for consistency between text, figures and tables, to provide feedback on manuscript language and structure and to assist rephrasing of short sections of text for clarity and to meet word count limits. All generated outputs were reviewed, tested and verified by the authors.

\section{Results and Discussion}
\label{sect:results}

\subsection{Final Models Inference}
\label{sect:model_inf_results}

The final WRF2WRF model was selected after 160 epochs, with validation-set pixelwise RMSE of 0.11, SSIM of 0.74, and PSNR of 19.56. These metrics were computed in the normalized color-encoded space rather than converted to wind speed, and are therefore dimensionless. For GOES2PARA, validation after 120 epochs yielded an RMSE of 0.05, SSIM of 0.94, and PSNR of 25.96. Less epochs were required for GOES2PARA to reach comparable image generation skill due to the larger training dataset and the structurally simpler parametric wind fields.

Inference was benchmarked on a CPU runtime on Google Colab with batch size = 1 to align with GOES real-time image availability. Per-image inference time averaged 255 ms for WRF2WRF and 266 ms for GOES2PARA. Full pipeline processing including image input, parameter conditioning, output and visualization averaged 630 ms and 357 ms respectively. Faster pipeline processing for GOES2PARA is attributed to the simplicity of the output wind fields compared to finer scale detail represented in WRF2WRF. Our parametric model (Section 2\ref{sect:para_methods}, Appendix~\ref{app:parametric}), by comparison, takes 9,400 ms for the full pipeline, 26 times slower than GOES2PARA. These short timings on CPU, combined with the hourly GOES data upload latency of $\le$1 minute \citep{NOAA_GOES_MRD_2024} suggests suitability for near real-time operational TC wind field diagnosis, with low compute requirements for both models.

\subsection{Model performance on Test Set}
\label{sect:res_model_test_set}

\begin{figure}[H]
  \centerline{\includegraphics[width=30pc]{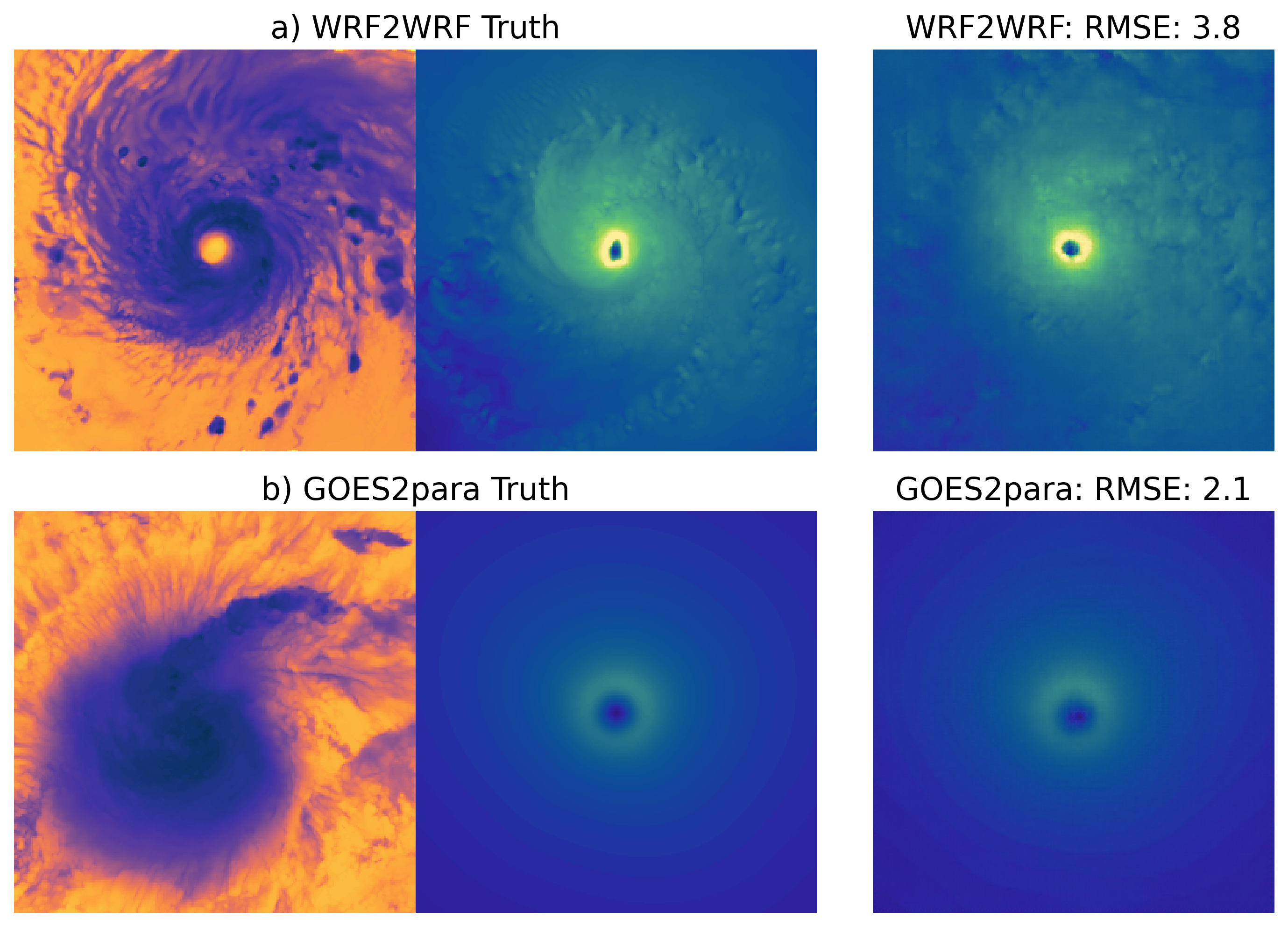}}
\caption{Example test set truth and prediction images with pixel-wise RMSE for (a) WRF2WRF Hurricane Lee 2023 b) GOES2PARA Hurricane Jerry 2019.}
  \label{test_set_images}
\end{figure}

WRF2WRF and GOES2PARA learned to construct wind fields from IR inputs, capturing the main features of the storm including eyewall, RMW, storm asymmetry and intensity across the wind field. In addition WRF2WRF generated more detailed eyewall and outer rainband structures. Example image results from both models are included in Fig.~\ref{test_set_images}.

\begin{figure}[H]
  \centerline{\includegraphics[width=30pc]{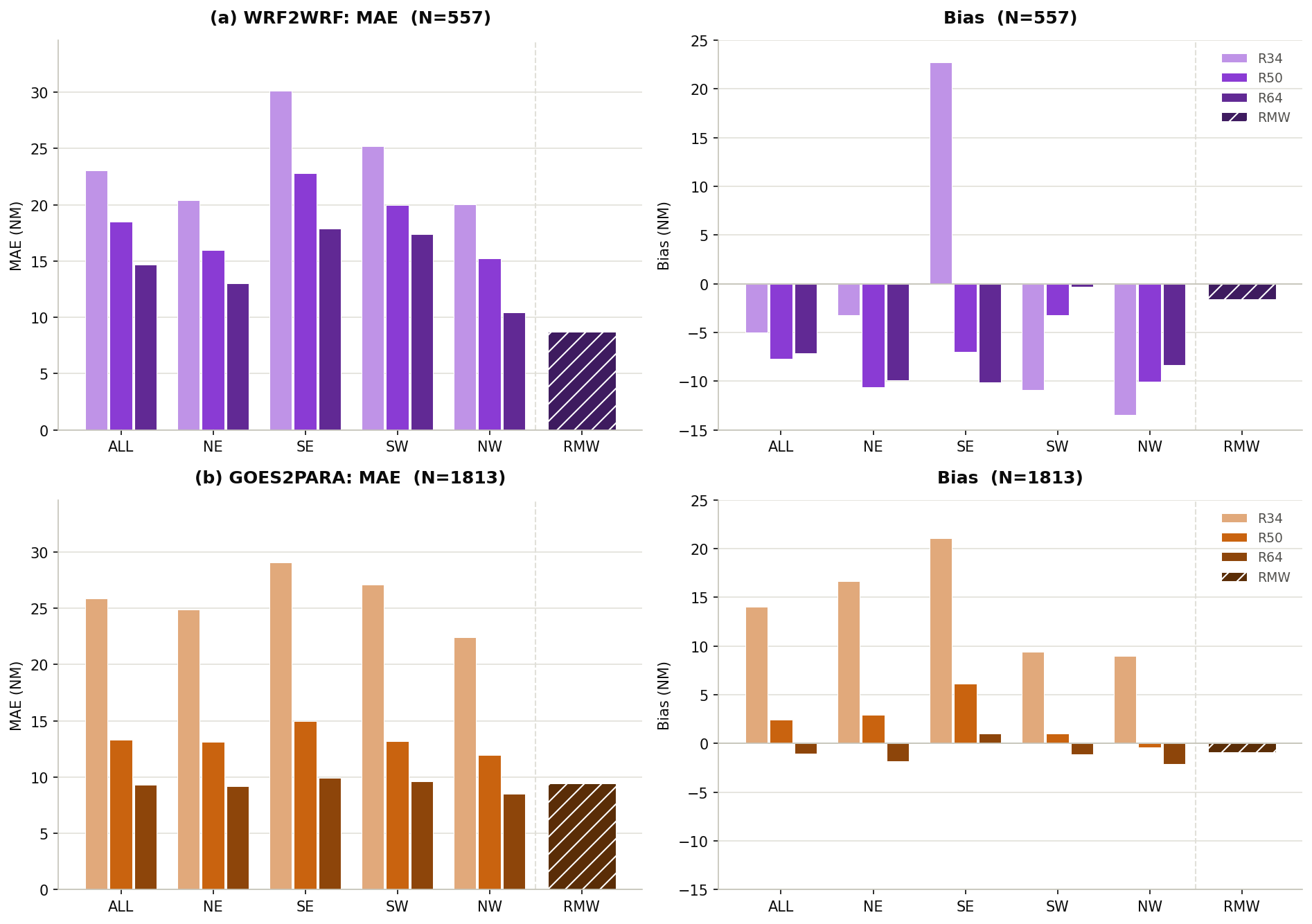}}
\caption{All radii, R34, R50, R64 and RMW Mean Absolute Error (MAE) and bias for inference on the whole test set for each model a)WRF2WRF (557 images), b)GOES2PARA (1813 images).}
  \label{mae_bias_model_truth}
\end{figure}

For WRF2WRF pixel-wise RMSE between model inference on the test set and ground truth was 7.2 kt with R34 MAE of 22.5 NM (bias $-0.1$ NM) for all quadrants (Fig.~\ref{mae_bias_model_truth}, Table ~\ref{tab:testset_gan_results}). The lowest WRF2WRF R34 MAE was in the NW quadrant (20 NM, bias -13.5, Fig.~\ref{mae_bias_model_truth}), though overall bias was considerably lower for R34 (-0.1 NM). For WRF2WRF, as expected, MAE improved for higher wind speed radii (R50, R64), with the lowest MAE for R64 NW quadrant (8.5 NM, bias -8.3 NM (Fig.~\ref{mae_bias_model_truth}). For WRF2WRF ablation studies, there was negligible improvement in pixel-wise RMSE for WRF2WRF(lat) and WRF2WRF(Vmax) compared to the full WRF2WRF model (Table~\ref{tab:testset_gan_results}). Whilst this pixel-wise wind speed reconstruction was relatively insensitive to the inclusion of scalar/vector conditioning parameters, RMW and wind radii estimation were sensitive to parameter choice. The full model performed best on RMW (7.8 NM, bias -2.0 NM), outperforming all ablation variants by at least 2.3 NM (Table~\ref{tab:testset_gan_results}). Among the single parameter ablations WRF2WRF(lat) and WRF2WRF(Vmax) outperformed other ablation variants on RMW (10.1 and 10.8 NM MAE, respectively, versus 11.1 NM for the base configuration). The improvement from latitude conditioning may reflect an indirect pathway, whereby latitude influences RMW through its effect on R34 \citep{Chavas_Knaff_2022_rmw}. Conditioning on ${V}_{\text{max}}$ improves RMW MAE with negligible bias (+0.5 NM), explained by the more direct physical dependence of RMW on ${V}_{\text{max}}$(\cite{Chavas_Knaff_2022_rmw}.

For wind radii, WRF2WRF(base) showed slight MAE improvement relative to the full model for R34 (21.3 NM and 22.5 NM, respectively) but with higher bias, whilst WRF2WRF(uv) outperformed the full model on R50 (17.7 NM, bias +1.2, Table ~\ref{tab:testset_gan_results}), had lower MAE and bias than WRF2WRF(base) at R64 (17.6 NM, -10.0 ) and maintained relatively low bias throughout. We interpret this as motion-vector conditioning contributing more to outer-wind-field structure while performing poorly for RMW, plausibly because $u$ and $v$ encode storm-relative asymmetry, whereas ${V}_{\text{max}}$ and $\phi$ conditioning contribute more to inner-core structure such as RMW.

GOES2PARA excelled at pixel-wise RMSE (6.4 kt). This was expected as predicting the parametric contoured wind field is a simpler generative task than the finer scale wind field perturbations predicted by WRF2WRF. GOES2PARA RMW performance (9.5 NM, bias - 0.9 NM) was not as strong as WRF2WRF, but it outperformed the WRF2WRF ablation models. GOES2PARA R34 MAE for all quadrants was 25.9 NM (bias + 14.1 NM; Table~\ref{tab:testset_gan_results}, Fig.~\ref{mae_bias_model_truth}), the highest R34 bias for all of the models. This bias, as for WRF2WRF, is dominated by SE quadrant bias. Again, quadrant-resolved results showed the lowest R34 MAE in the NW quadrant (22.4 NM, bias 9.0 NM, Fig.~\ref{mae_bias_model_truth}). For GOES2PARA the lowest R34 MAE was in NW quadrant (8.5 NM, bias -2.1 NM) as with WRF2WRF MAE and bias improves with higher wind speed radii (R50, R64), Table ~\ref{tab:testset_gan_results}, Fig.~\ref{mae_bias_model_truth}).  As with WRF2WRF, MAE and bias improved at higher wind radii (R50, R64; Table~\ref{tab:testset_gan_results}) and GOES2PARA achieved the lowest R64 MAE of any of the models (9.3 NM, -1.0 bias).

\begin{table*}[h!]
\begin{center}
\caption{Pix2pix GAN test set performance for WRF2WRF and GOES2PARA models with Mean Absolute Error (MAE) and bias values for all images.}
\label{tab:testset_gan_results}
\small
\setlength{\tabcolsep}{3pt}
\renewcommand{\arraystretch}{1.1}
\begin{tabular}{@{}lcccccc@{}}
\topline
\textbf{Parameters} & WRF2WRF & \makecell{WRF2WRF\\(base)} & \makecell{WRF2WRF\\(lat)} & \makecell{WRF2WRF\\(vmax)} & \makecell{WRF2WRF\\(uv)} & GOES2PARA \\
\midline
pixel-wise RMSE (kt) & 7.2 & 7.3 & 7.1 & 7.0 & 7.3 & 6.4 \\
RMW MAE (Bias) (NM) & 7.8 (-2.0) & 11.1 (-0.6) & 10.1 (-2.5) & 10.8 (+0.5) & 13.9 (+0.3) & 9.5 (-0.9) \\
R34 MAE (Bias) (NM) & 22.5 (-0.1) & 21.3 (-4.3) & 27.9 (+5.3) & 21.8 (-2.7) & 23.4 (+4.2) & 25.9 (+14.1) \\
R50 MAE (Bias) (NM) & 18.5 (-7.8) & 17.2 (-5.5) & 18.96 (-8.5) & 18.0 (-9.2) & 17.7 (+1.2) & 13.3 (+2.4) \\
R64 MAE (Bias) (NM) & 14.7 (-7.1) & 17.6 (-10.0) & 16.4 (-8.0) & 16.2 (-12.2) & 16.4 (-3.5) & 9.3 (-1.0) \\
\bottomrule
\end{tabular}
\end{center}

\end{table*}

\subsection{Comparison with Operational Best Track}
\label{sect:res_model_vs_ibtracs}

\begin{figure}[H]
  \centerline{\includegraphics[width=30pc]{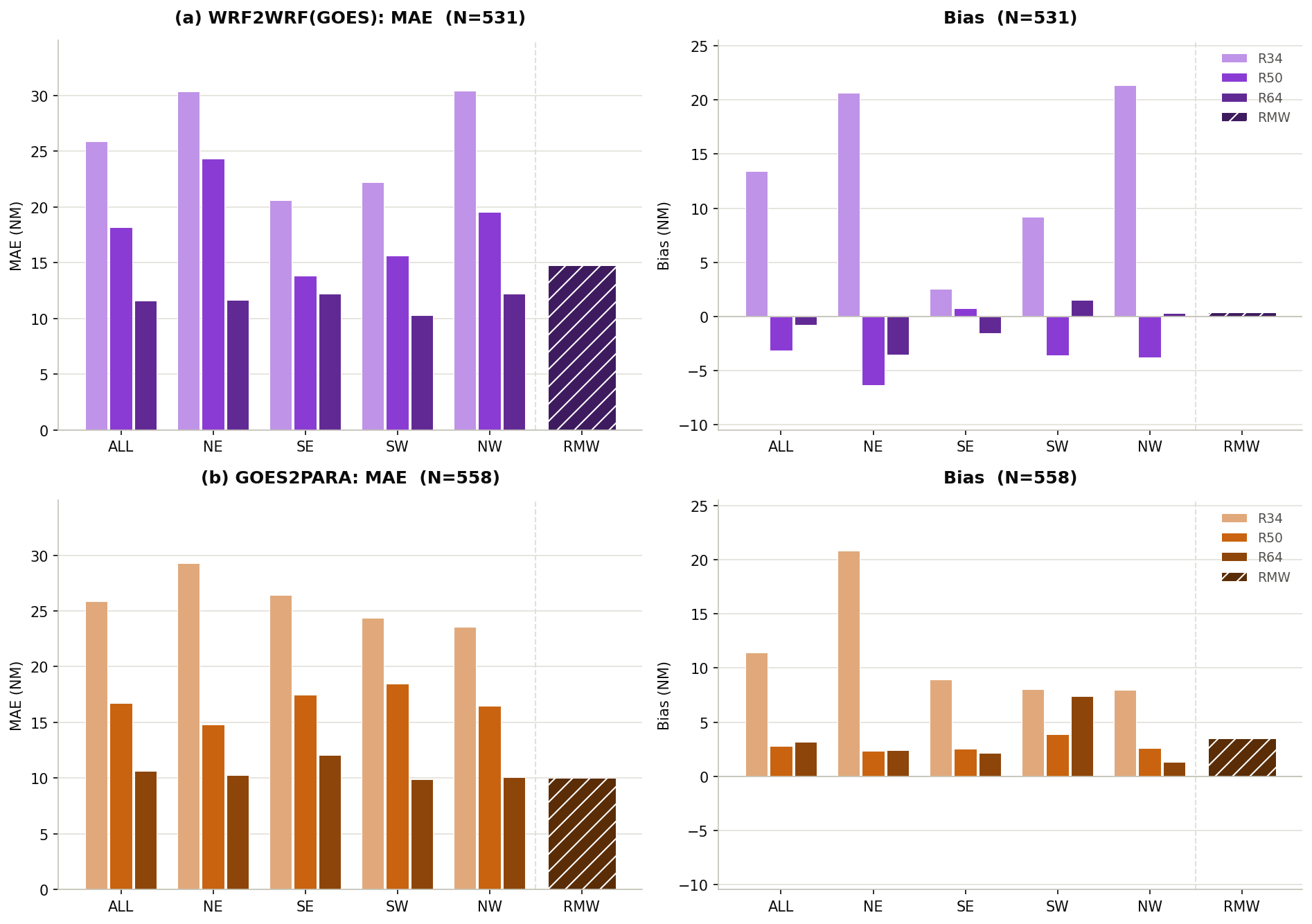}}
\caption {R34, R50, R64 and RMW Mean Absolute Error (MAE) and bias for all radii and by quadrant for WRF2WRF and GOES2PARA inference with GOES IR compared with ibtracs radii.}
  \label{fig: models_vs_ibtracs_radii}
\end{figure}

For all comparisons with operational best track (IBTrACS) per quadrant bias was calculated between model derived and IBTrACS radii for each radii threshold (R34, R50, R64) on the validation set. We then used this validation set bias to bias correct the model-derived test set radii before comparison with IBTrACS. When benchmarked against operational best track (IBTrACS) both models resolved RMW well with MAE 14.8 NM (bias + 0.4 NM) and 10.0 NM (bias + 3.5 NM) for WRF2WRF(GOES) and GOES2PARA, respectively (Fig.~\ref{fig: models_vs_ibtracs_radii}, Table~\ref{tab:testset_gan_results_ibtracs}). As expected, absolute errors decreased with increasing wind threshold for both models (Table~\ref{tab:testset_gan_results_ibtracs}). Whilst WRF2WRF(GOES) resolved R34 (MAE 25.9 NM, bias + 13.4 NM) comparably to GOES2PARA (25.9 NM, bias +11.4 NM), GOES2PARA outperformed WRF2WRF(GOES) for R50 (16.8 NM, bias + 2.8 NM, 18.2 NM, bias –3.1 NM, respectively) and R64 (10.6 NM, bias + 3.2 NM and 11.6, bias –0.8 NM, respectively), Fig.~\ref{fig: models_vs_ibtracs_radii}, Table~\ref{tab:testset_gan_results_ibtracs}). The degraded performance of WRF2WRF(GOES) compared to GOES2PARA, is expected as WRF2WRF(GOES) was subjected to a more challenging evaluation. GOES2PARA was evaluated with test data from the same distribution as its training and validation data. In contrast, WRF2WRF(GOES) evaluates WRF2WRF, trained exclusively on simulated data, with a domain shift to real GOES IR with different characteristics to simulated IR. Furthermore, WRF2WRF(GOES) evaluates WRF2WRF on a larger and more varied dataset than the WRF2WRF validation set that was used to select the final WRF2WRF model.

Table \ref{tab:mae_comparison} places the mean absolute error (MAE) of WRF2WRF(GOES) and GOES2PARA in the context of previously published wind radii estimates from statistical \citep{Kossin2007, Knaff2016, Dolling2016}, and deep learning methodologies \citep{ZhuoTan2021, Wang2023}. Both models perform comparably to, and for some quadrants exceed, the published benchmarks, though direct comparison is complicated by the heterogeneity in datasets between studies that span different time periods and basins. 

GOES2PARA's R64 errors (9.9-12.1 NM) are broadly comparable to the deep learning benchmarks range of 8.8-9.5 NM \citep{ZhuoTan2021, Wang2023}, despite GOES2PARA data covering a longer time period (2004–2025), though we note \citet{Wang2023} evaluated on a larger storm sample (21 TCs). At R50, GOES2PARA (MAE 14.8–18.5 NM) outperforms the statistical baselines \citep{Kossin2007, Knaff2016} in all quadrants and is comparable to \citet{Dolling2016}, better in NE and NW, worse in SE and SW. Whilst GOES2PARA generalizes reasonably well across a wide range of storm intensities and structures, it remains around 30–70\% less accurate than the deep learning benchmarks. R34 performance is weak (23.6–29.3 NM) compared to the Deep Learning studies (17.0–19.4 NM \citet{ZhuoTan2021, Wang2023}, with an improvement over \citet{Dolling2016} in the NW quadrant.

WRF2WRF(GOES) shows a similar improvement towards higher wind thresholds (20.6–30.4 NM at R34, 10.3–12.2 NM at R64). However, its R34 errors are markedly asymmetric: the SE and SW quadrants (20.6 and 22.2 NM) are competitive with \citet{Dolling2016} and outperform \citet{Kossin2007}, but the NE and NW quadrants (both 30.4 NM) are comparable to or worse than all statistical baselines and roughly 60\% less accurate than the deep learning methods (17.0–19.4 NM). WRF2WRF(GOES) R50 NE error (24.4 NM) exceeds all published benchmarks, including the statistical baselines, whereas its R50 SE and SW errors (13.8 and 15.7 NM) remain competitive with \citet{Dolling2016}. At R64 WRF2WRF(GOES) errors are more uniform (10.3–12.2 NM), with the SW and NW quadrants still above the best published values (6.9–9.1 NM in \citet{Dolling2016}). The performance of WRF2WRF(GOES), provides promising evidence that the WRF2WRF transfer learning step preserves the learned wind field reasonably well when adapting to observational GOES target data. 

In Fig.~\ref{fig:ibtracs_r34 comparison} both model predictions tend to overestimate R34 when best-track radii are small and underestimate R34 when best-track radii are larger, a relationship that was also observed for North Atlantic storms in \citet{Wang2023}. GOES2PARA, which is able to predict radii up to 210 NM compared to 154.5 NM for WRF2WRF(GOES), benefits from its training wind field truth data being constructed from IBTrACS radii and has higher correlation with IBTrACS radii (R = 0.69 - 0.75) than WRF2WRF(GOES) (R = 0.47 - 0.63). Whilst GOES2PARA outperformed WRF2WRF(GOES) at for R64, WRF2WRF(GOES) performed best for SE and SW quadrants for R50 (Appendix~\ref{app:r50r64_ibtracs}). Both models show slightly lower correlations (R = 0.47–0.75) than \citet{Wang2023} (R = 0.83–0.88) and slightly larger R34 MAEs (20.6–30.4 NM versus 17.0–21.0 NM). 

The reduction in performance for the GAN models in this study compared to the CNNs in \citet{Wang2023} may be explained by two factors: the larger training dataset used for the CNNs allowing for more robust feature learning, and the comparatively simpler task of direct radii estimation \citep{Wang2023} relative to the current study's task, which requires complete wind field generation from which wind radii are then calculated. Whilst underestimation of radii above 140 NM is consistent across quadrants in \citet{Wang2023}, GOES2PARA and WRF2WRF(GOES) show less underestimation in NE and SE and NE and NW, respectively. \citet{Wang2023} attributes over- and underestimation at the TC radius at upper and lower TC size extremes to unbalanced TC wind radii data distributions; whilst the current data have a similar imbalance it appears less pronounced than in \citet{Wang2023}.

\begin{figure}[H]
    \centerline{\includegraphics[width=40pc]{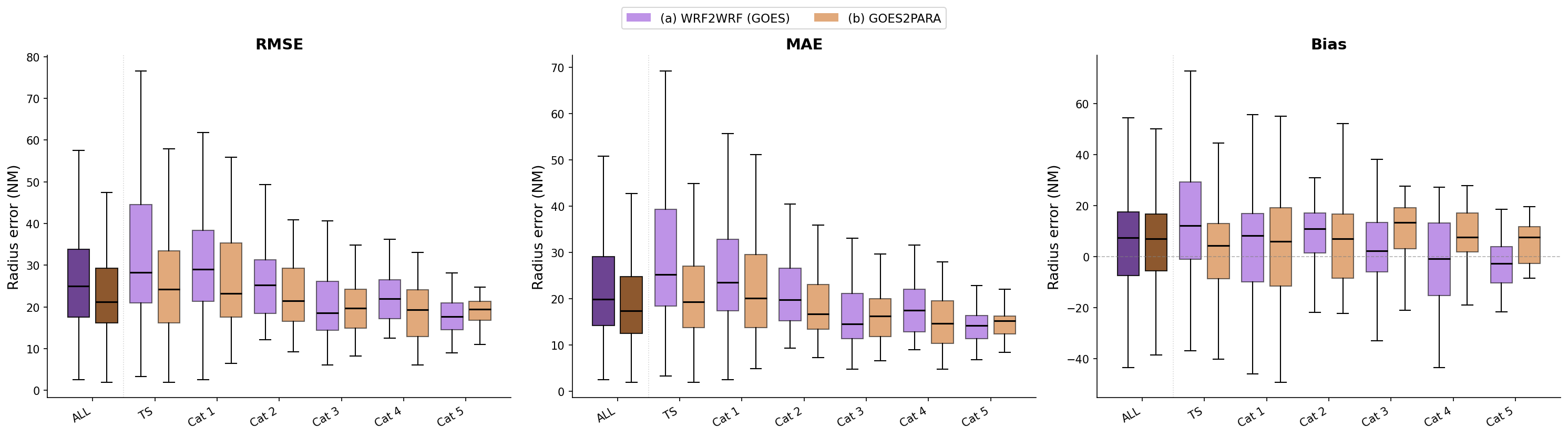}}
    \caption{Mean Absolute Error (MAE) and bias for all radii by storm category for WRF2WRF (purple), GOES2PARA (orange) inference on GOES data, compared with IBTrACS.}
    \label{fig:radii_by_stormCat}
\end{figure}

\begin{figure}[H]
    \centerline{\includegraphics[width=40pc]{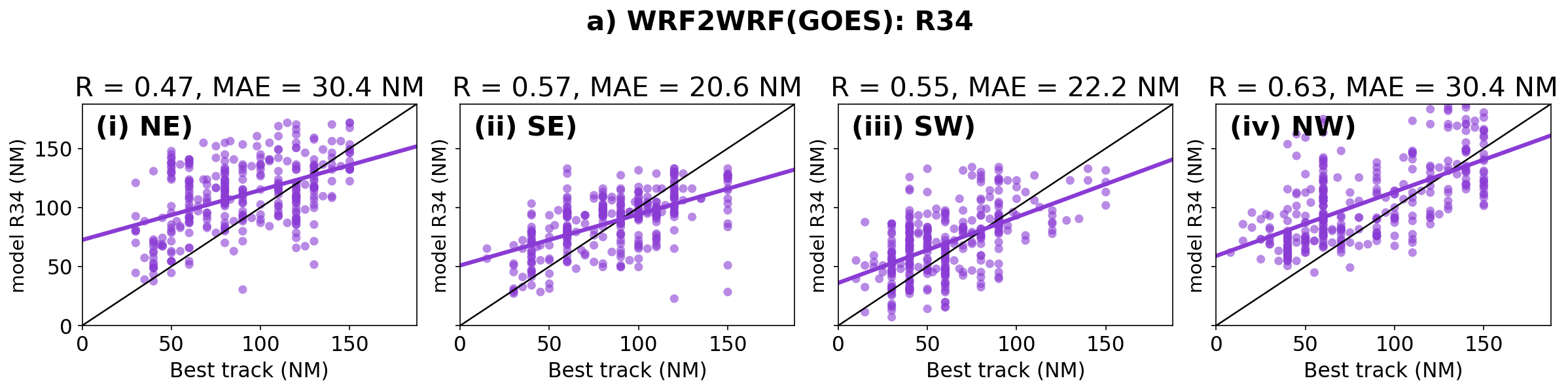}}
    \centerline{\includegraphics[width=40pc]{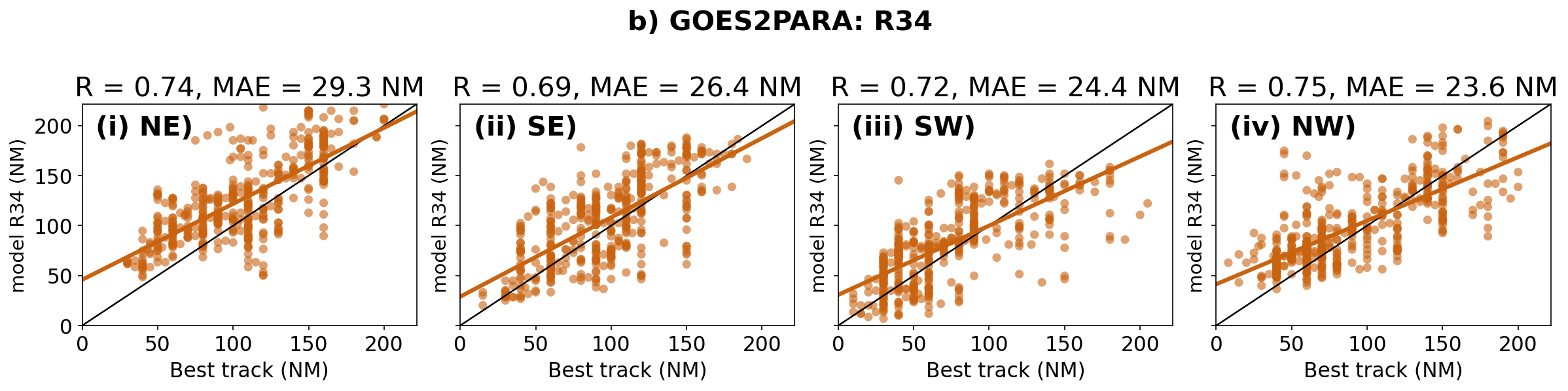}}
    \caption{R34 comparison with IBTrACS vs mpdel inference on GOES input IR images for a) WRF2WRF(GOES) b) GOES2PARA for quadrants i) NE, ii) SE, iii) S, iv) NW. R values for the regression lines are listed in plot headers alongside MAE. }
    \label{fig:ibtracs_r34 comparison}
\end{figure}

Category-resolved MAE (Fig.~\ref{fig:radii_by_stormCat}) reveals modest storm-intensity dependence in radii estimation skill. WRF2WRF(GOES) and GOES2PARA exhibit the lowest MAE for Category 5 storms relative to best-track. These higher Category storms are more organized, providing structured patterns for the models to learn and infer; improved model skill at higher storm intensities has also been reported in recent deep learning studies \citep{Ollier_2025, Wimmers2024}. However, even the most intense storms spend relatively little time at Category 5, so in our dataset this storm category is sparsely represented in both test sets: 83 images (4.6\%) for GOES2PARA and 84 images (5.3\%) for WRF2WRF(GOES). Given these small sample sizes, the apparent skill advantage at Category 5 should be interpreted with caution. Whilst \citet{Wimmers2024} evaluate flight-level radial winds and RMW against aircraft data rather than surface wind radii against IBTrACS; they likewise cautioned that the apparent skill improvement for the highest category storms is limited by small high-intensity samples.

We note a caveat applies to the interpretation of the GOES2PARA best-track comparison. As GOES2PARA is trained on wind fields that the parametric model constructs from best-track parameters (RMW and quadrant R34/R50/R64, Section 2\ref{sect:ibtracs_methods}, Appendix~\ref{app:parametric}), benchmarking against IBTrACS (Table 4, Figs. 6–7) does not provide a fully independent validation. However, at model inference GOES2PARA receives only single-channel IR and three scalar/vector parameters and reconstructs the full wind field without access to the target radii data. Thus agreement with IBTrACS is non-trivial and reflects translation from the IR input. Section 3\ref{sect:res_model_vs_SAR} provides a fully independent validation.

\begin{table}
\caption{Pix2pix GAN GOES test set performance for WRF2WRF and GOES2PARA models against IBTrACS truth. MAE (Bias) (NM), by quadrant.}
\label{tab:testset_gan_results_ibtracs}
\footnotesize
\begin{center}
\begin{tabular}{lccccc}
\topline
\textbf{Model} & \textbf{ALL} & \textbf{NE} & \textbf{SE} & \textbf{SW} & \textbf{NW}\\
\midline
\textit{RMW} &  &  &  &  &  \\
WRF2WRF(GOES)   & 14.8 (+ 0.4) & -- & -- & -- & -- \\
GOES2PARA & 10.0 (+ 3.5) & -- & -- & -- & -- \\
\midline
\textit{R34} &  &  &  &  &  \\
WRF2WRF(GOES)   & 25.9 (+ 13.4) & 30.4 (+ 20.6) & 20.6 (+ 2.5) & 22.2 (+ 9.2) & 30.4 (+ 21.4)\\
GOES2PARA & 25.9 (+ 11.4) & 29.3 (+ 20.9) & 26.4 (+ 9.0)  & 24.4 (+ 8.0) & 23.6 (+ 8.0) \\
\midline
\textit{R50} &  &  &  &  &  \\
WRF2WRF(GOES)   & 18.2 (-- 3.1) & 24.4 (-- 6.3) & 13.8 (+ 0.8) & 15.7 (-- 3.6)  & 19.6 (-- 3.8)\\
GOES2PARA & 16.8 (+ 2.8)  & 14.8 (+ 2.3)   & 17.5 (+ 2.6)  & 18.5 (+ 3.9) & 16.5 (+ 2.6)\\
\midline
\textit{R64} &  &  &  &  &  \\
WRF2WRF(GOES)   & 11.6 (-- 0.8) & 11.7 (-- 3.5) & 12.2 (-- 1.5) & 10.3 (+ 1.5) & 12.2 (+ 0.3)  \\
GOES2PARA & 10.6 (+ 3.2)  & 10.3 (+ 2.4) & 12.1 (+ 2.1)  & 9.9 (+ 7.4)  & 10.1 (+ 1.3)  \\
\botline
\end{tabular}
\end{center}
\end{table}

\begin{table*}[t!]
\caption{Comparison of R34, R50 and R64 MAE (NM) by quadrant for WRF2WRF(GOES) and GOES2PARA models and models in published literature. Table adapted from \citet{Wang2023}.}
\label{tab:mae_comparison}
\begin{center}
\footnotesize
\setlength{\tabcolsep}{3pt}
\begin{tabular}{lp{2.6cm}cccccccccccc}
\topline
& & \multicolumn{4}{c}{\textbf{R34}} & \multicolumn{4}{c}{\textbf{R50}} & \multicolumn{4}{c}{\textbf{R64}} \\
\textbf{Method} & \textbf{Test data} & \textbf{NE} & \textbf{SE} & \textbf{SW} & \textbf{NW} & \textbf{NE} & \textbf{SE} & \textbf{SW} & \textbf{NW} & \textbf{NE} & \textbf{SE} & \textbf{SW} & \textbf{NW}\\
\midline
\textbf{WRF2WRF}(GOES) & 7 TCs 2008--2025 & 30.4 & 20.6 & 22.2 & 30.4 & 24.4 & 13.8 & 15.7 & 19.6 & 11.7 & 12.2 & 10.3 & 12.2\\
\textbf{GOES2PARA} & 8 TCs 2004--2025 & 29.3 & 26.4 & 24.4 & 23.6 & 14.8 & 17.5 & 18.5 & 16.5 & 10.3 & 12.1 & 9.9 & 10.1 \\
\midline
\citet{Kossin2007} & 2005 & \multicolumn{4}{c}{24.2} & \multicolumn{4}{c}{19.8} & \multicolumn{4}{c}{14.5}\\
\citet{Knaff2016} & 2011--13 & 44 & 37 & 36 & 41 & 21 & 21 & 23 & 23 & 13 & 13 & 13 & 14\\
\citet{Dolling2016} & 21 TCs 2004--2010 & 27.7 & 25.2 & 19.9 & 30.1 & 17.4 & 16.7 & 12.3 & 18.1 & 10.3 & 8.9 & 6.9 & 9.1\\
\citet{ZhuoTan2021} & 2017 and 2019 & \multicolumn{4}{c}{17.0} & \multicolumn{4}{c}{12.5} & \multicolumn{4}{c}{9.2}\\
\citet{Wang2023} & 21 TCs 2004--2010 & 18.8 & 17.8 & 17.5 & 19.4 & 11.5 & 11.2 & 10.9 & 10.9 & 8.9 & 9.5 & 9.0 & 8.8\\
\botline
\end{tabular}
\end{center}
\end{table*}

\subsection{Comparison with SAR Data}
\label{sect:res_model_vs_SAR}

WRF2WRF(GOES) and GOES2PARA were further evaluated against concurrent SAR wind fields for test-set storms (42 SAR scenes for WRF2WRF(GOES), 28 for GOES2PARA). Because SAR wind fields do not share any information with the training targets, this evaluation provides an independent assessment of wind-field skill. GOES2PARA yielded a mean pixel-wise RMSE of 12.1 kt and MAE of 9.5 kt, slightly lower than WRF2WRF(GOES) (RMSE 13.9 kt, MAE 11.3 kt),  GOES2PARA also had a lower MAE (9.4 kt) and bias (5.3 kt) than WRF2WRF (MAE 11.3 kt, bias 5.9 kt) (Fig.~\ref{fig:box_model_sar_compare}). As expected, the parametric model outperformed both GOES2PARA and WRF2WRF on RMSE (9.5 kt), MAE (7.5 kt) and bias (3.2 kt). Since the parametric model requires IBTraCS RMW and wind radii inputs, which are unavailable in near real-time operational settings, it cannot provide the valuable diagnostic information that the GANs can. The similarity in bias values for GOES2PARA and the parametric model are consistent with GOES2PARA learning the inherent bias of the parametric dataset. That GOES2PARA and WRF2WRF(GOES) were 
only 2 - 4 kt less accurate than the parametric model, despite lacking access to its storm structure inputs, offers promising evidence that these diagnostic GAN models can recover useful structural information from IR observations earlier than operational best-track analysis becomes available.

\begin{figure}[h!]

  \centerline{\includegraphics[width=30pc]{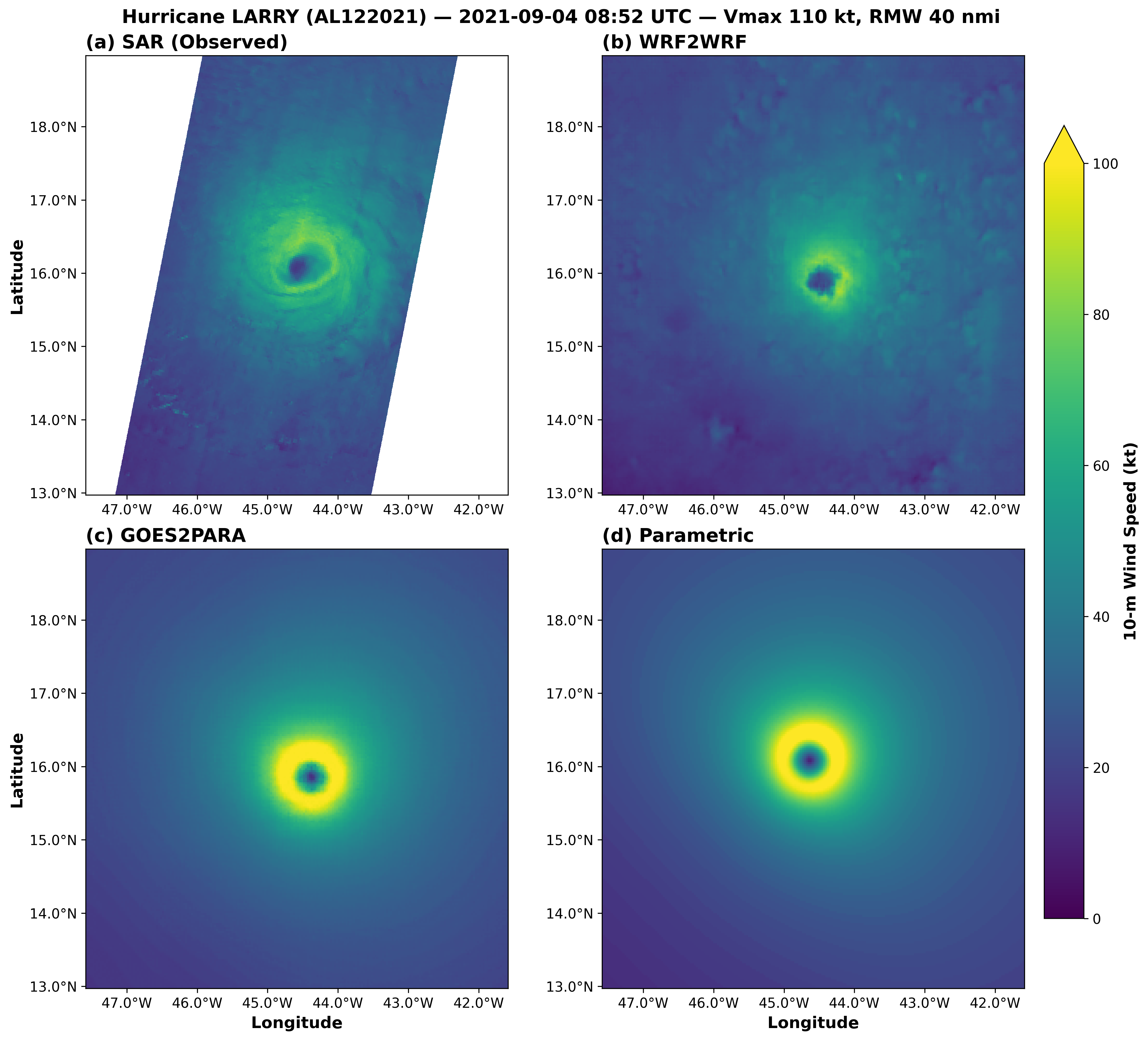}}
\caption{Example model to SAR comparison for Hurricane Larry with GOES data as input IR. for a)SAR (b) WRF2WRF(GOES) (c) GOES2PARA (d) the parametric model. Note the SAR image displayed is at WRF2WRF spatial extent (6$^{\circ}\times$6$^{\circ}$), whilst GOES2PARA and the parametric model predictions are based on  7$^{\circ}\times$7$^{\circ}$ spatial extent of the same image.}

  \label{fig:img_model_sar_compare}
\end{figure}

\begin{figure}[h!]

  \centerline{\includegraphics[width=35pc]{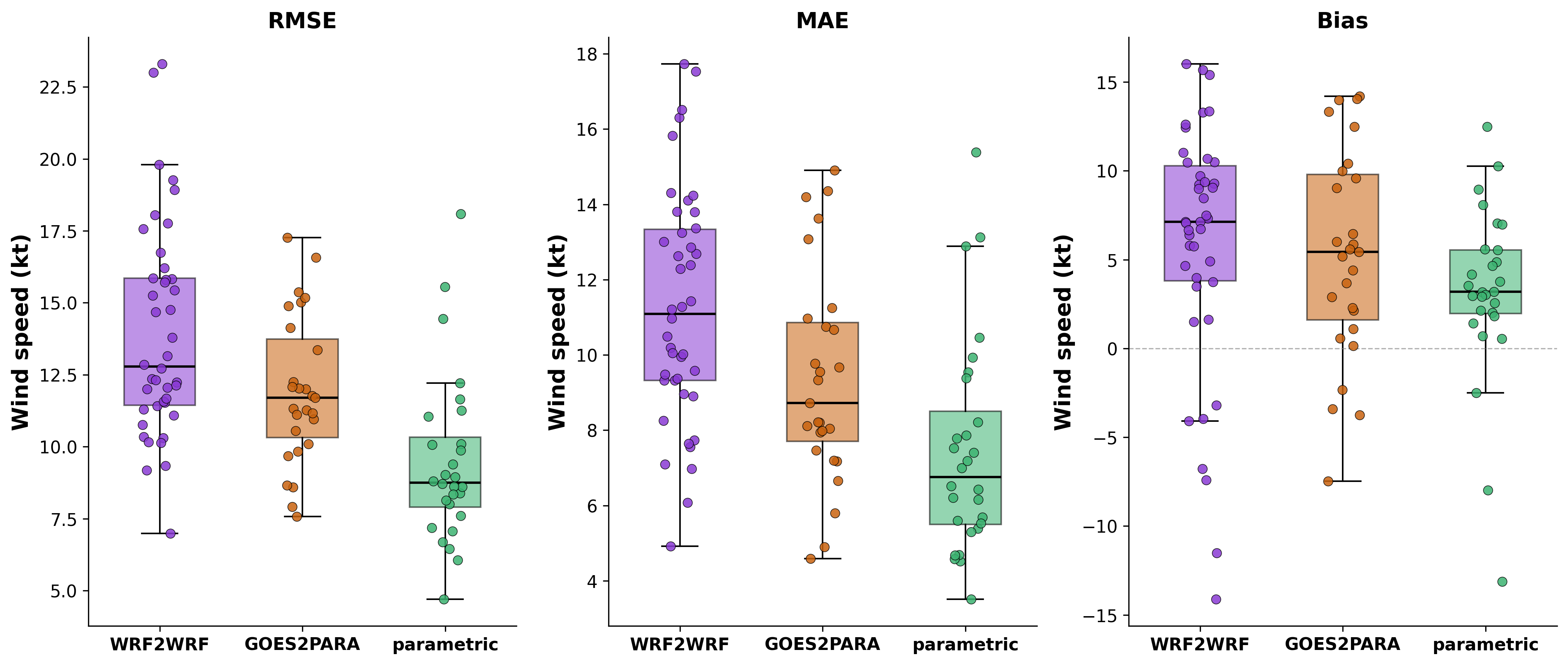}}
\caption{pixel-wise RMSE, MAE and Bias for GOES inference compared with SAR images for WRF2WRF(GOES) (purple) and GOES2PARA (orange). The parametric model (green) is included for reference.}
  \label{fig:box_model_sar_compare}
\end{figure}

Example model-to-SAR comparisons for Hurricane Larry 2021 are shown in Fig.~\ref{fig:img_model_sar_compare}. For this example WRF2WRF pixel-wise RMSE was 7.0 kt, with MAE 4.9 (bias -3.2), close to GOES2PARA performance of RMSE 8.6 kt, with MAE and bias of 4.6 and -2.9, respectively. Direct whole wind field comparisons with published work are limited, as few IR-to-wind-field studies validate against SAR. The closest analogue is \citet{Chen2025}, which reported a $V_{max}$ MAE of 11.9 kt against 92 independent SAR scenes from 2017--2018 storms. Following their methodology GOES2PARA achieved 12.6 kt $V_{max}$ MAE demonstrating a similar skill level. The pixel-wise validation presented above, however, is a stricter test, as it additionally penalizes errors in wind-field asymmetry.

\section{Conclusions and Future Work}
\label{sect:conclusions}

This study addresses data availability gaps in high resolution Tropical Cyclone (TC) wind fields due to the scarcity of direct wind field measurements during active storms. We developed two parameter conditioned image-to-image satellite IR wind field diagnosis models that generate whole storm two-dimensional wind fields. Unlike existing statistical and deep learning models for TCs, which require multiple data inputs for training and inference, our WRF2WRF and GOES2PARA models require a single channel IR plus three scalars/vectors: coriolis proxy $sin(\phi)$,  $V_{max}$ and storm motion vectors ($u, v$) that are readily available in operational advisories. Both models produce high temporal-spatial resolution wind fields rapidly with low compute and input data requirements. Furthermore, the low per-image inference times of 255–266 ms on a single CPU core, combined with GOES upload latency of $\leq$1 minute has the potential for near-real-time operational TC diagnosis, providing an updated whole storm wind field with every new satellite image at a fraction of the cost of the parametric model (357 ms versus 9,400 ms full-pipeline).

GOES2PARA, trained on real IR satellite images paired with parametric wind fields that are built on the extended best-track data set of TC intensities and wind radii (54 storms, 11,886 pairs), resolves RMW and the asymmetries of the outer wind field allowing for diagnosis of R64, R50, and R34 wind radii in each storm quadrant. WRF2WRF, trained on IR to wind mapping from realistic simulations of TC IR and wind fields for 24 storms (3,323 images), provides the same wind-radii information but additionally reproduces finer-scale structures associated with rain bands and eye wall asymmetries. 

WRF2WRF(GOES) tested WRF2WRF on GOES IR imagery, a challenging zero shot domain transfer from simulated to real IR imagery on a much larger heterogeneous dataset than was used to train WRF2WRF. WRF2WRF(GOES) transferred well to the GOES IR imagery with similar, and in some quadrants better, performance to GOES2PARA in wind radii estimation when benchmarked against operational best-track data. WRF2WRF achieved mean absolute errors of 14.8 nautical miles (NM) for radius of maximum wind (RMW) and 25.9 NM for 34 knot wind radii. GOES2PARA achieved mean absolute errors of 10.0 NM for RMW and 25.9 nautical miles for 34-knot wind radii. 

Both WRF2WRF(GOES) and GOES2PARA were further evaluated against Synthetic Aperture Radar wind fields, which share no information with the GOES2PARA training targets. GOES2PARA yielded a lower mean pixel-wise error (RMSE 12.1 kt, MAE 9.5 kt) than WRF2WRF (RMSE 13.0 kt, MAE 10.5 kt). The parametric model achieved the lowest error on this sample (RMSE 9.5 kt, MAE 7.5 kt), but at roughly 26 times the per-image cost of GOES2PARA, highlighting the speed advantage of the GAN approaches. The close agreement in bias between GOES2PARA and the parametric model is consistent with GOES2PARA having learned the inherent bias of the parametric model.

Currently GOES2PARA and WRF2WRF are optimized for offshore TCs, future models will incorporate roughness length into the parametric model for GOES2PARA to account for land effects and introduce land falling TCs into the WRF2WRF dataset. In addition, both models are optimized for North Atlantic storms, currently the models do not account for differences in storm physical characteristics in other storm basins. With two different sets of training data but the same model architecture and similar hyper-parameters, both models yielded comparable results; this provides the framework for transferring the models to other storm basins. As each storm basin has unique storm characteristics, training basin specific models using the same methodologies with basin specific data should yield the best results per basin. 

GOES2PARA and WRF2WRF are limited to past and current TC diagnosis, however, the presented methodology is adaptable to short-term prognosis. Pairing current wind field with IR imagery from 3, 6 or 9 hours prior provided promising results in preliminary tests. Pairing current wind field with current IR and prior IR would also provide additional storm evolution context with the potential to improve wind field diagnosis.

\clearpage

\acknowledgments
Research funding was provided by Worldsphere INC. Compute resources were provided by Worldsphere INC and University of Miami Frost Institute for Data Science and Computing. All Authors declare part ownership of Worldsphere INC.

\datastatement
GOES ABI IR imagery was obtained from the NOAA Registry of Open Data on AWS (https://registry.opendata.aws/noaa-goes/). GRIDSAT-B1 IR data were obtained from NOAA NCEI (https://www.ncei.noaa.gov/products/gridded-geostationary-brightness-temperature). IBTrACS best track data is available from NOAA NCEI (https://www.ncei.noaa.gov/products/international-best-track-archive). All the WRF simulations presented here are archived at the University of Miami Institute for Data Science and Computing and are available for remote transfer. The parametric model data are archived at Worldsphere INC and are available for remote transfer.

\newcommand{\thd}[1]{\parbox[b]{\linewidth}{\bfseries\raggedright #1\strut}}
\newcommand{\thdc}[1]{\parbox[b]{\linewidth}{\bfseries\centering #1\strut}}

\appendix[A]
\appendixtitle{WRF simulations}

\makeatletter\edef\@currentlabel{A}\label{app:wrf}\makeatother

\footnotesize
\renewcommand{\arraystretch}{1.1}
\setlength{\tabcolsep}{2.5pt}
\setlength{\LTleft}{0pt}
\setlength{\LTright}{0pt}

\begin{longtable}{
  >{\raggedright\arraybackslash}p{0.100\textwidth}
  >{\centering\arraybackslash}p{0.055\textwidth}
  >{\raggedright\arraybackslash}p{0.100\textwidth}
  >{\raggedright\arraybackslash}p{0.125\textwidth}
  >{\centering\arraybackslash}p{0.055\textwidth}
  >{\raggedright\arraybackslash}p{0.110\textwidth}
  >{\centering\arraybackslash}p{0.075\textwidth}
  >{\centering\arraybackslash}p{0.075\textwidth}
  >{\raggedright\arraybackslash}p{0.135\textwidth}
}
\caption{Summary of WRF tropical cyclone simulations, including domain
configuration, initial/boundary conditions, peak intensity at time of peak,
and mean simulated peak wind and mean track latitude over the full
simulation.}\label{tab:wrf_simulations}\\
\topline
\thd{Storm/\newline simulation} &
\thdc{WRF\newline version} &
\thd{Region} &
\thd{Grid sizes} &
\thdc{Hourly\newline outputs} &
\thd{Peak intensity\newline \& latitude} &
\thdc{Mean peak\newline wind (kt)} &
\thdc{Mean center\newline latitude ($^\circ$N)} &
\thd{Initial\slash Boundary\newline conditions} \\
\midline
\endfirsthead

\caption[]{Summary of WRF tropical cyclone simulations (continued).}\\
\topline
\thd{Storm/\newline simulation} &
\thdc{WRF\newline version} &
\thd{Region} &
\thd{Grid sizes} &
\thdc{Hourly\newline outputs} &
\thd{Peak intensity\newline \& latitude} &
\thdc{Mean peak\newline wind (kt)} &
\thdc{Mean center\newline latitude ($^\circ$N)} &
\thd{Initial\slash Boundary\newline conditions} \\
\midline
\endhead

HNR1 & 3.1.1 & Tropical Atlantic &
27: 240$\times$160\newline 9: 120$\times$120\newline 3: 240$\times$240\newline 1: 480$\times$480 &
240 & 122 kt at 23.4N & 87.0 & 24.0 & Joint OSSE Nature Run\textsuperscript{1} \\[6pt]

HNR2 & 3.4.1 & Caribbean\slash Gulf of Mexico &
9: 480$\times$360\newline 3: 360$\times$360\newline 1: 480$\times$480 &
48 & 80 kt at 26.7N & 57.0 & 23.4 & Joint OSSE Nature Run \\[6pt]

Bill (2009) & 3.4.1 & Western Atlantic &
9: 384$\times$336\newline 3: 360$\times$360\newline 1: 480$\times$480 &
79 & 133 kt at 27.1N & 122.6 & 21.8 & GFDL analysis \\[6pt]

Ideal3 & 3.4.1 & Zonally periodic channel\textsuperscript{2} &
27: 240$\times$180\newline 9: 180$\times$180\newline 3: 360$\times$360\newline 1: 480$\times$480 &
216 & 115 kt at 20.0N & 83.1 & 20.7 & Idealized vortex with mean flow and shear \\[6pt]

Ideal4\textsuperscript{3} & 3.4.1 & Zonally periodic channel &
27: 240$\times$180\newline 9: 180$\times$180\newline 3: 360$\times$360\newline 1: 480$\times$480 &
107 & 125 kt at 20.0N & 88.2 & 20.4 & Idealized vortex with mean flow and shear \\[6pt]

Ideal5 & 3.4.1 & Zonally periodic channel &
27: 240$\times$180\newline 9: 180$\times$180\newline 3: 360$\times$360\newline 1: 480$\times$480 &
216 & 149 kt at 15.0N & 111.4 & 15.3 & Idealized vortex with mean flow and shear \\[6pt]

Wilma\textsuperscript{4} (2005) & 3.9.1.1 & Gulf of Mexico &
9: 288$\times$256\newline 3: 240$\times$240\newline 1: 432$\times$432 &
34 & 114 kt at 24.9N & 105.5 & 24.9 & GFS analysis \\[6pt]

Maria\textsuperscript{5} (2017) & 3.9.1.1 & Eastern Caribbean &
9: 336$\times$288\newline 3: 240$\times$240\newline 1: 432$\times$432 &
67 & 138 kt at 17.1N & 118.6 & 17.6 & GFS analysis \\[6pt]

Beryl (2025) & 4.3.3 & Caribbean &
27: 240$\times$155\newline 9: 180$\times$180\newline 3: 240$\times$240\newline 1: 360$\times$360 &
112 & 117 kt at 16.5N & 91.4 & 14.2 & ERA5 \\[6pt]

Erin (2025) & 4.3.3 & West Atlantic &
27: 180$\times$180\newline 9: 180$\times$180\newline 3: 240$\times$240\newline 1: 360$\times$360 &
145 & 116 kt at 24.2N & 89.0 & 25.7 & ERA5 \\[6pt]

Ernesto (2024) & 4.3.3 & West Atlantic &
27: 180$\times$180\newline 9: 180$\times$180\newline 3: 240$\times$240\newline 1: 360$\times$360 &
133 & 100 kt at 28.3N & 71.2 & 27.7 & ERA5 \\[6pt]

Florence (2018) & 4.3.3 & West Atlantic &
27: 180$\times$120\newline 9: 120$\times$120\newline 3: 240$\times$240\newline 1: 480$\times$480 &
156 & 129 kt at 28.9N & 96.2 & 27.1 & ERA5 \\[6pt]

Franklin (2023) & 4.3.3 & West Atlantic &
27: 180$\times$180\newline 9: 180$\times$180\newline 3: 240$\times$240\newline 1: 360$\times$360 &
157 & 125 kt at 29.2N & 84.2 & 27.3 & ERA5 \\[6pt]

Gabrielle (2025) & 4.3.3 & West Atlantic &
27: 180$\times$180\newline 9: 180$\times$180\newline 3: 240$\times$240\newline 1: 360$\times$360 &
126 & 97 kt at 39.5N & 65.4 & 32.2 & ERA5 \\[6pt]

Helene (2024) & 4.3.3 & Gulf of Mexico &
27: 180$\times$180\newline 9: 180$\times$180\newline 3: 240$\times$240\newline 1: 360$\times$360 &
50 & 85 kt at 26.8N & 64.7 & 22.5 & ERA5 \\[6pt]

Humberto (2025) & 4.3.3 & West Atlantic &
27: 180$\times$180\newline 9: 180$\times$180\newline 3: 240$\times$240\newline 1: 360$\times$360 &
133 & 125 kt at 28.6N & 81.9 & 24.0 & ERA5 \\[6pt]

Kirk (2024) & 4.3.3 & Central Atlantic &
27: 180$\times$180\newline 9: 180$\times$180\newline 3: 240$\times$240\newline 1: 360$\times$360 &
133 & 115 kt at 27.5N & 83.9 & 23.4 & ERA5 \\[6pt]

Lee (2023) & 4.3.3 & Central Atlantic &
27: 180$\times$180\newline 9: 180$\times$180\newline 3: 240$\times$240\newline 1: 360$\times$360 &
205 & 118 kt at 21.2N & 95.1 & 24.3 & ERA5 \\[6pt]

Margot (2023) & 4.3.3 & Central Atlantic &
27: 180$\times$180\newline 9: 180$\times$180\newline 3: 240$\times$240\newline 1: 360$\times$360 &
205 & 86 kt at 28.7N & 64.8 & 29.0 & ERA5 \\[6pt]

Melissa (2025) & 4.3.3 & Caribbean &
27: 180$\times$180\newline 9: 180$\times$180\newline 3: 240$\times$240\newline 1: 360$\times$360 &
109 & 120 kt at 19.7N & 71.9 & 17.0 & ERA5 \\[6pt]

Philippe (2023) & 4.3.3 & Central Atlantic &
27: 180$\times$180\newline 9: 180$\times$180\newline 3: 240$\times$240\newline 1: 360$\times$360 &
239 & 122 kt at 30.6N & 70.7 & 22.0 & ERA5 \\[6pt]

Tammy (2023) & 4.3.3 & Central Atlantic &
27: 180$\times$180\newline 9: 180$\times$180\newline 3: 240$\times$240\newline 1: 360$\times$360 &
99 & 111 kt at 25.9N & 70.6 & 21.1 & ERA5 \\[6pt]

Teddy (2020) & 4.3.3 & Central Atlantic &
27: 180$\times$180\newline 9: 180$\times$180\newline 3: 240$\times$240\newline 1: 360$\times$360 &
175 & 121 kt at 38.3N & 81.9 & 22.9 & ERA5 \\
\botline
\end{longtable}

\footnotesize
\noindent\textsuperscript{1}The Joint OSSE Nature Run is described in
\citet{Andersson2010}.\\
\textsuperscript{2}The zonal channel simulations use the f-plane framework so
the Coriolis parameter and the effective latitude are constant across the
domain.\\
\textsuperscript{3}Ideal4 was not used in \citet{Klotz_Nolan_2019}, but is
nearly identical to Ideal3, except that the central SST was increased from
27$^\circ$C to 28$^\circ$C.\\
\textsuperscript{4}The Wilma simulation used here is case ``FinalYSU'' as
described in Nolan et al. (2021).\\
\textsuperscript{5}The Maria simulation is described in \citet{Ge2019}.
\normalsize
\appendix[B]
\appendixtitle{Parametric model}
\makeatletter\edef\@currentlabel{B}\label{app:parametric}\makeatother

The parametric model used for this study utilizes a modified-Rankine vortex wind profile and readily-available parameters in the best-track data. The input parameters are the intensity ($V_{\text{max}}$), radius of maximum wind ($RMW$), four-quadrant radius of tropical storm force wind ($R34$), and the TC position (latitude and longitude). The storm's forward speed and heading are calculated from those track coordinates.

\begin{figure}[H]
  \centerline{\includegraphics[width=29pc]{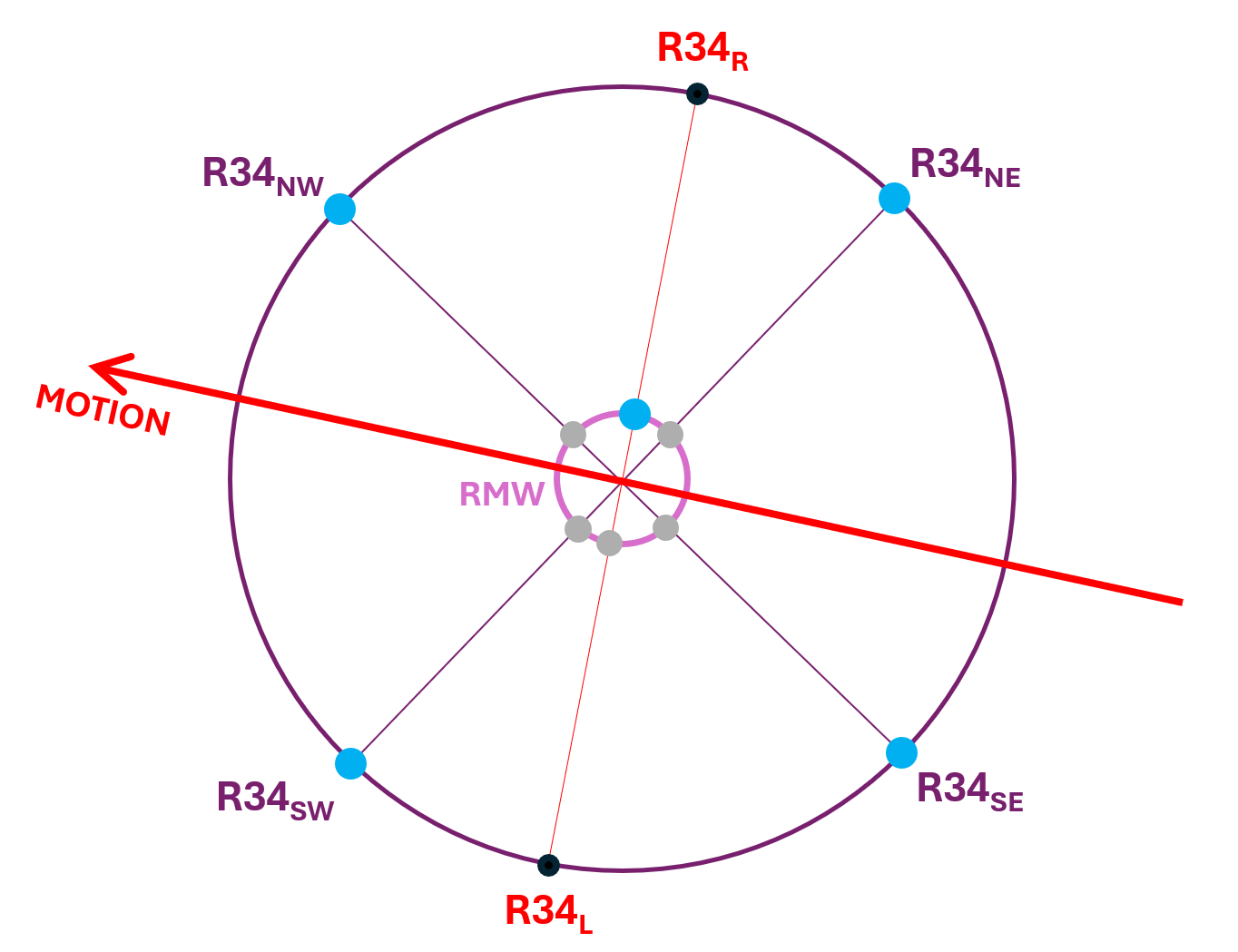}}
\caption{Schematic diagram of the azimuthal control points relative to the four quadrants and storm motion. Values at the blue dots, as well as the RMW and storm motion, come from the best-track data. The values of wind speed at the RMW at the gray dots are calculated from the translation asymmetry at five known azimuths, and values at the black dots are calculated using a Hermite spline.}
  \label{parametric_schematic}
\end{figure}

This model is based upon defining azimuthal control points, which are locations around the storm center where one or more parameters are known. We create six of these points: four of them correspond to the cardinal directions associated with the northeast, southeast, southwest, and northwest quadrants with $R34$ values; the other two are assigned to the perpendicular directions right and left of storm motion. Those six azimuths and their locations relative to the storm center are illustrated in Figure~\ref{parametric_schematic}.

The vortex is created on a polar grid with radial spacing of 4 km and azimuthal spacing of 2$^{\circ}$.  This is adequate to maintain the details in structure and asymmetry and produce a satisfactory final wind field.

We use the translation asymmetry factor described in \cite{Knaff2007} (K07) which is dependent on storm speed, $c$, and latitude, $\phi$:
\begin{equation}
a = a_0 + a_1*c + a_2*c^2 + a_3*(\phi-25) 
\label{eq_para1}
\end{equation}
where $a_0$ = 1.06, $a_1$ = 0.28, $a_2$ = -0.0026, $a_3$ = -0.08, $c$ is in knots, and $\phi$ is in degrees.

At each of the six azimuthal control points, $V_{\text{max}}$ is adjusted by the asymmetry parameter such that it maintains the best-track value at the point 90$^{\circ}$ to the right of storm motion ($\theta_R$) and is a minimum at the point 90$^{\circ}$ left of the storm motion ($\theta_L$). At each of the control points, $n$,
\begin{equation}
v_n = (V_{\text{max}}-a) + (a*cos(\theta_n-\theta_R))
\end{equation}
where $\theta_n$ is one of the six control point azimuths and $\theta_R$ is the angle 90$^{\circ}$ to the right of storm motion (based on Eqn 1 in K07). Note that in K07, the asymmetry is applied to the entire wind field, while in this model it is only applied to the wind speed at the $RMW$ because the best-track $R34$ values already account for the translation asymmetry. Finally, a zero-derivative Hermite spline is fit to all six $v_n$ values so there are intensity values, $v$, at each 2$^{\circ}$ azimuth in the polar grid. That particular spline was chosen as it smoothly connects the data points without over- or undershooting them, maintaining the exact best-track values.

Next, $R34$ is already available at four of the six control points ($R34_{NE}$, $R34_{SE}$, $R34_{SW}$, $R34_{NW}$), but is also needed at $\theta_R$ and $\theta_L$. To calculate those, a zero-derivative Hermite spline is fit to the existing $R34s$, and the values located at $\theta_R$ and $\theta_L$ are used for $R34_R$ and $R34_L$.

The final component to creating the radial wind profile is the Rankine decay exponent, $\alpha$:
\begin{equation}
\alpha_n = |ln(v_n/34)/ln(R34_n/RMW)|   
\end{equation}
To help alleviate occasional numerical problems with weaker tropical tropical storms ($< \sim\!50\ kt$) and/or asymmetric storms where $R34$ is close to $RMW$ creating unrealistic $\alpha$, $\alpha$ has some constraints placed on it through the use of size and intensity-based default decay behavior:
\begin{equation}
\alpha_0 = 0.6 + 0.2*EXP(-RMW/40).\\
\end{equation}
That parameter, when needed, assigns a default decay exponent of 0.6 and modifies it such that it decays quicker for smaller $RMW$ and slower for larger $RMW$.
Then we define a weighting function
\begin{equation}
\omega = 1 - EXP(-(dv/v_{trans})^2)
\end{equation}
where $dv = v_n-34$ (how far above tropical storm intensity) and $v_{trans} = 15$ (transition zone to apply the weighting). Assigning the original $\alpha_n$ values to $\alpha_{orig}$, the updated $\alpha_n$ values are
\begin{equation}
\alpha_n = \omega*\alpha_{orig} + (1-\omega)*\alpha_0.
\end{equation}
For stronger tropical storms and hurricanes, this weighting function adjustment is never used. As before, a zero-derivative Hermite spline is fit to those updated six $\alpha_n$ values, so at each 2$^{\circ}$ azimuth, we now have smoothly-evolving and well-behaved values of $RMW$, $v$, $R34$, and $\alpha$.

Now the modified-Rankine vortex equations are applied at each of the 2$^{\circ}$ azimuths from r = 0-1000 km from the storm center:
\begin{equation}
v_{outer} = v * (RMW/r)^\alpha
\end{equation}
\begin{equation}
v_{inner} = v * r/RMW
\end{equation}

We remove the undesirable wind speed cusp at $RMW$ while maintaining the peak value using
\begin{equation}
v = w*v_{inner} + (1-w)*v_{outer},
\end{equation}
where $w = 1 / (1+(r/RMW)^k)$ and $k$ is a sharpness factor set to 6 here (higher values keep more of the cusp while lower values remove more of the cusp).

\begin{figure}
  \centerline{\includegraphics[width=29pc]{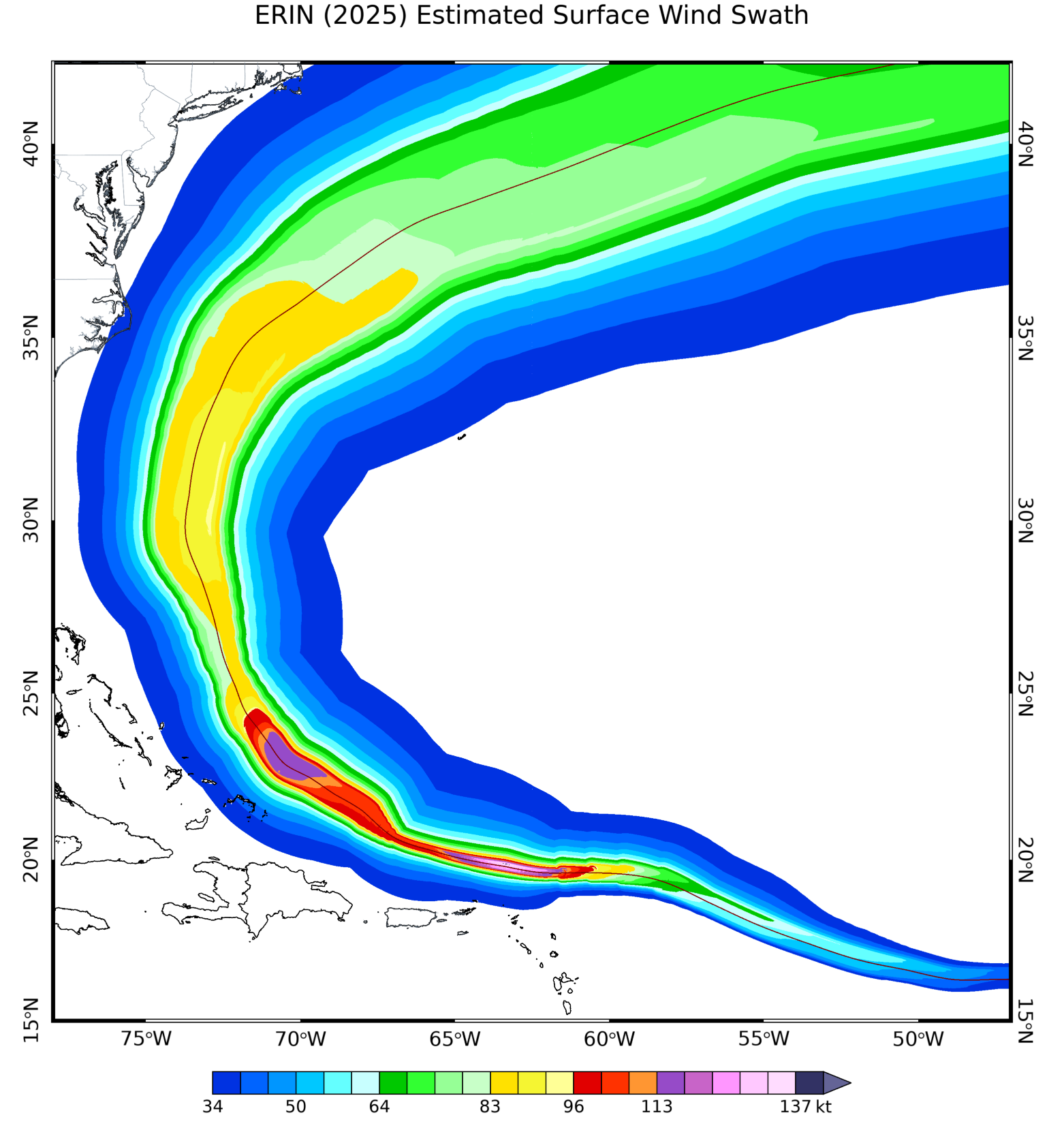}}
\caption{A wind swath created for Hurricane Erin (2025). This is the result of accumulating approximately 3200 individual parametric vorticies along the track interpolated to five-minute time steps. Color families in the contouring correspond to Saffir-Simpson wind scale categories, and the contour intervals are distributed evenly within a category (approximately 5 kt intervals).}
  \label{windswath_example}
\end{figure}

That final expression for $v$ is calculated everywhere on the polar grid to arrive at the full 2D surface wind field.  The work described in this paper uses hourly snapshots of these wind fields, but an additional useful application of this model is to interpolate the best-track data to 5-minute increments and integrate it over the time steps to produce a smooth wind swath, an example of which from Hurricane Erin 2025 is shown in Figure~\ref{windswath_example}.

\appendix[C]
\appendixtitle{R50 and R64 comparison with operational best track}

\makeatletter\edef\@currentlabel{C}\label{app:r50r64_ibtracs}\makeatother

\begin{figure}
    \centerline{\includegraphics[width=30pc]{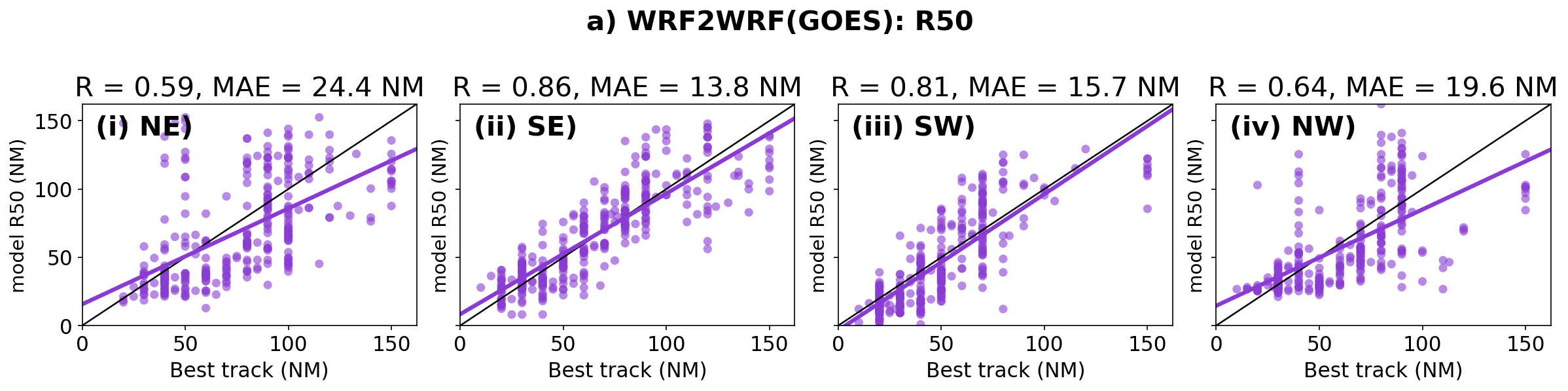}}
    \centerline{\includegraphics[width=30pc]{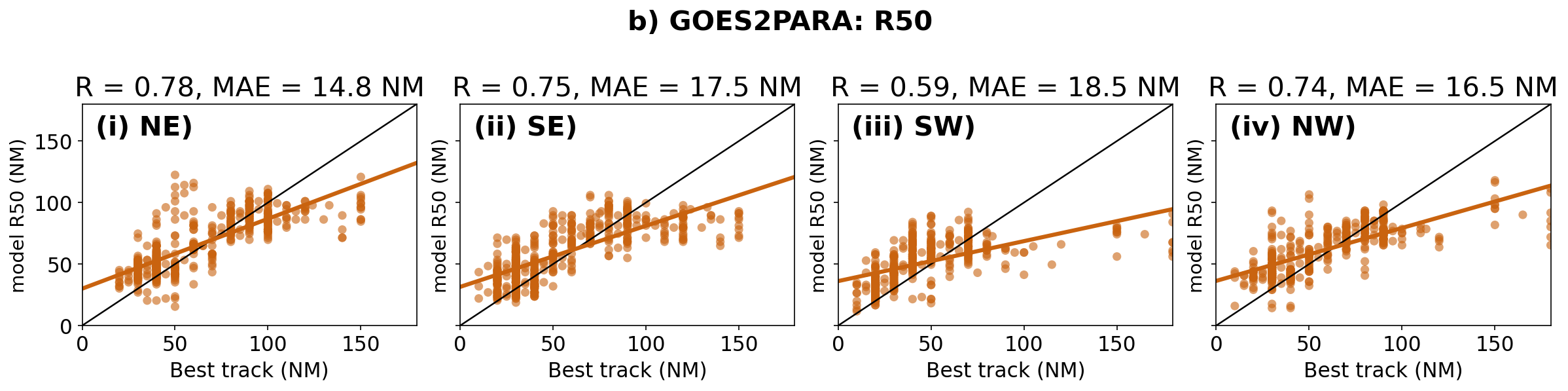}}
    \caption{R50 comparison with IBTrACS vs model inference on GOES input IR images for (a) WRF2WRF(GOES) (b) GOES2PARA or quadrants i) NE, ii) SE, iii) S, iv) NW. R values for the regression lines are listed in plot headers alongside MAE.}
    \label{fig:ibtracs_r50 comparison}
\end{figure}

\begin{figure}
    \centerline{\includegraphics[width=30pc]{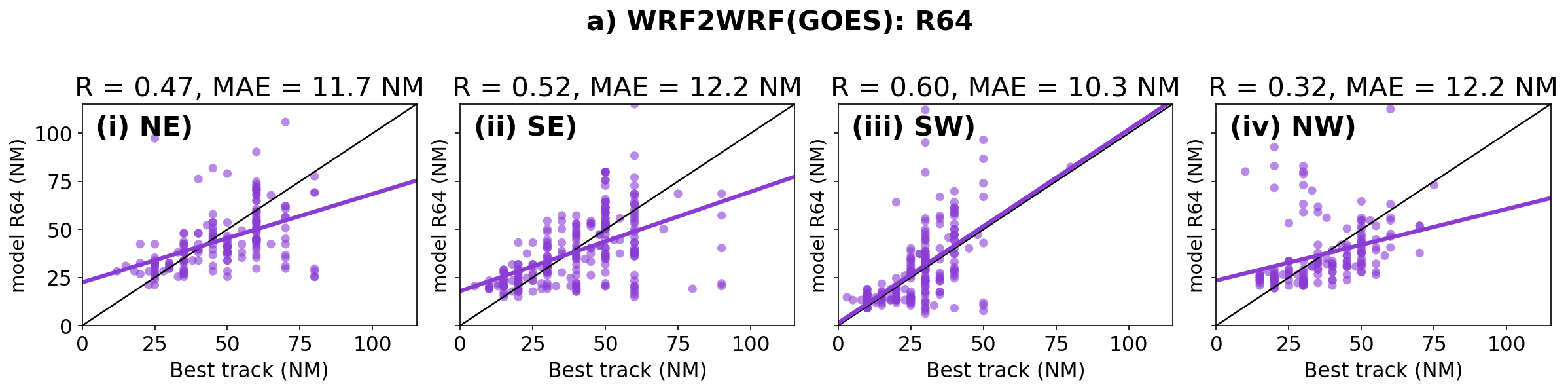}}
    \centerline{\includegraphics[width=30pc]{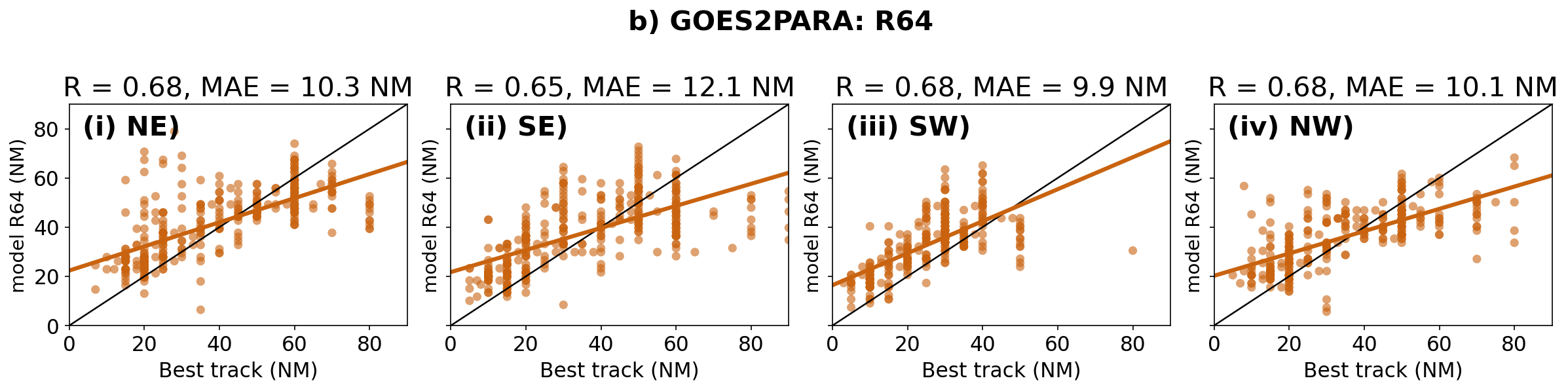}}
    \caption{As ~\ref{fig:ibtracs_r50 comparison} for R64}
    \label{fig:ibtracs_r64 comparison}
\end{figure}

Fig.~\ref{fig:ibtracs_r50 comparison} and Fig.~\ref{fig:ibtracs_r64 comparison} show the correlation between model and best track for R50 and R64, respectively for WRF2WRF and GOES2PARA. As with R34 (Fig.~\ref{fig:ibtracs_r34 comparison}, Section 3\ref{sect:res_model_vs_ibtracs}), for both models there is overestimation for smaller radii and underestimation for larger radii, with the exeption of WRF2WRF(GOES) SW quadrants. Again, this is largely in agreement with North Atlantic storms in \citet{Wang2023}.


\bibliographystyle{ametsocV6}
\bibliography{references}

\end{document}